%% file: main.tex
\documentclass[sigconf,nonacm]{acmart}
\AtBeginDocument{%
  }

\setcopyright{acmlicensed}
\copyrightyear{2018}
\acmYear{2018}
\acmDOI{XXXXXXX.XXXXXXX}

\newcommand{\new}[1]{\textcolor{black}{#1}}

\usepackage{quoting}
\usepackage{multicol}
\usepackage{multirow}
\usepackage{arydshln}
\usepackage{pifont}% 
\usepackage{array}
\usepackage{ragged2e}
\usepackage{listings}
\input{Figures/listing_style}

\usepackage{pgffor}   % for \foreach in \spec
\usepackage{makecell} % for \makecell in headers
\usepackage{booktabs}
\usepackage{array}
\usepackage[utf8]{inputenc}
\usepackage{enumitem}
\usepackage{xcolor}

\newcommand{\spec}[2]{%
  \foreach \i in {1,...,#2}{%
    \ifnum\i>#1 $\circ$\else $\bullet$\fi
  }%
}
\quotingsetup{font=itshape}

\begin{document}

%%
%% The "title" command has an optional parameter,
%% allowing the author to define a "short title" to be used in page headers.
\title[Behavior Latticing: Inferring User Motivations from Unstructured Interactions]{Behavior Latticing: \\Inferring User Motivations from Unstructured Interactions}

%%
%% The "author" command and its associated commands are used to define
%% the authors and their affiliations.
%% Of note is the shared affiliation of the first two authors, and the
%% "authornote" and "authornotemark" commands
%% used to denote shared contribution to the research.
\author{Dora Zhao}
\affiliation{%
  \institution{Stanford University}
  \city{Stanford}
  \state{CA}
  \country{USA}
  }
\email{dorothyz@stanford.edu}

\author{Michelle S. Lam}
\affiliation{%
  \institution{Stanford University}
  \city{Stanford}
  \state{CA}
  \country{USA}
}
\email{mlam4@cs.stanford.edu}

\author{Diyi Yang}
\affiliation{%
  \institution{Stanford University}
  \city{Stanford}
  \state{CA}
  \country{USA}
  }
\email{diyiy@cs.stanford.edu}

\author{Michael S. Bernstein}
\affiliation{%
  \institution{Stanford University}
  \city{Stanford}
  \state{CA}
  \country{USA}
  }
\email{msb@cs.stanford.edu}

\renewcommand{\shortauthors}{Zhao et al.}
\newcommand{\xmark}{\ding{55}}
\newcommand{\cmark}{\ding{51}}
\newcommand{\sysname}{\textsc{Dawn} }
\newcommand{\arc}{behavior latticing }
\newcommand{\arcnospace}{behavior latticing}
\newcommand{\pnum}{12 }
\newcommand{\tnum}{nine }
\newcommand{\tratings}{174 }
\newcommand{\sysnamenospace}{\textsc{Dawn}}
\newcommand{\techours}[0]{\textsc{Insights}}
\newcommand{\techbase}[0]{\textsc{Observations}}
\newcommand{\eebase}[0]{\textsc{Baseline}}
\definecolor{labelgray}{HTML}{383838}

\newcommand{\blacksquarelabel}[1]{%
  \begingroup
  \setlength{\fboxsep}{1.75pt}% tighter padding
  \raisebox{0.25ex}{% subtle vertical alignment tweak
    \colorbox{labelgray}{%
      \textcolor{white}{\footnotesize\makebox[0.9em][c]{#1}}%
    }%
  }%
  \endgroup
}
%%
%% The abstract is a short summary of the work to be presented in the
%% article.
\begin{abstract}
\input{Sections/00_Abstract_v3}
\end{abstract}

%%
%% The code below is generated by the tool at http://dl.acm.org/ccs.cfm.
%% Please copy and paste the code instead of the example below.
%%


%%
%% Keywords. The author(s) should pick words that accurately describe
%% the work being presented. Separate the keywords with commas.
%% A "teaser" image appears between the author and affiliation
%% information and the body of the document, and typically spans the
%% page.
\begin{teaserfigure}
  \includegraphics[width=\textwidth]{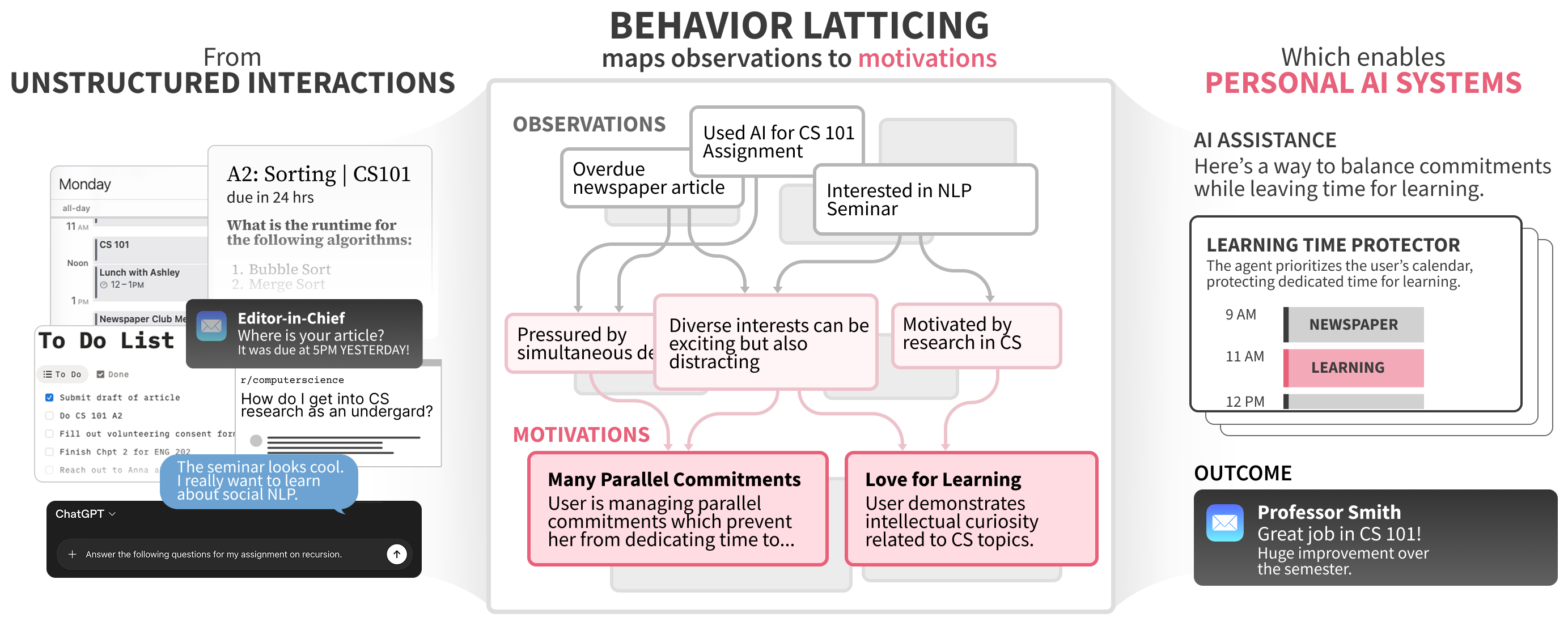}
  \caption{Today's personal AI systems focus on observations about what users do without considering why, thus constraining AIs to myopic task completion. In this work, we introduce \emph{behavior latticing}, an architecture for inferring insights about the motivations behind user behavior from unstructured interaction data. These insights enable the design of personal AI systems that can address users' underlying needs rather than only solving the task at hand.}
  \label{fig:teaser}
\end{teaserfigure}
\raggedbottom
\maketitle

\section{Introduction}
\input{Sections/01_Intro_v6}
\section{Related Work}
\begin{figure*}
    \centering
    \includegraphics[width=\linewidth]{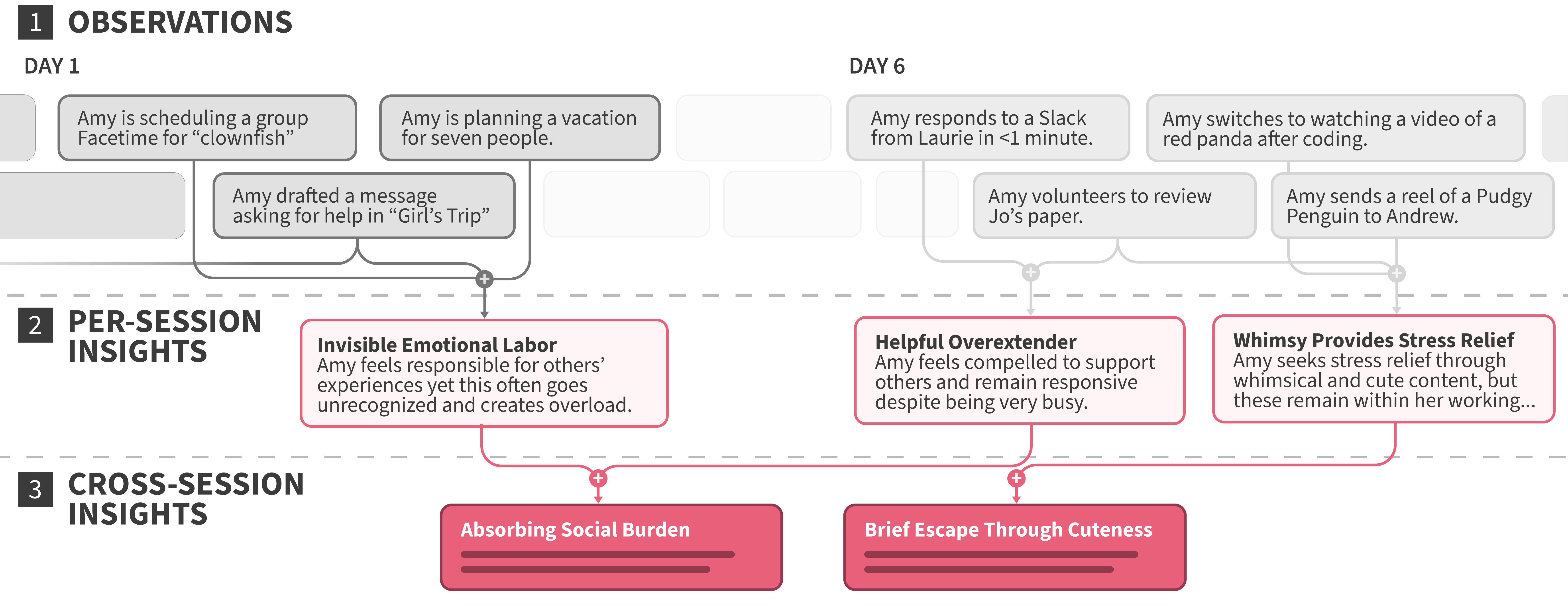}
    \caption{Our architecture synthesizes user insights through \emph{behavior latticing}. Starting from observations, we connect observations to produce insights within a single session (e.g., a conversation thread, a day of computer use). Each observation can be mapped to many insights. We recursively apply this step, leading to more cross-cutting inferences about the user.}
    \label{fig:insight_pipeline}
\end{figure*}
\input{Sections/02_RelatedWork}

\section{Understanding Users via Behavior Latticing}
\input{Sections/03_Architecture_v1}

\section{\sysnamenospace: Insight-Steered Personal AI Agents}
\input{Sections/04_Dawn}

\section{Evaluation Setup}
\label{sec:eval}
\input{Sections/05_Eval}
\section{Evaluating User Insights}
\label{sec:technical_eval}
\input{Sections/06_TechnicalEval}
\section{Evaluating \sysnamenospace}
\label{sec:e2e_eval}
\input{Figures/ComparingActions}
\input{Sections/07_ActionEval}
\section{Discussion}
\input{Sections/08_Discussion_v2}

\section{Conclusion}
\input{Sections/09_Conclusion}

\begin{acks}
We thank Yutong Zhang, Omar Shaikh, Jordan Troutman, Lindsay Popowski, Andrew Wang, and the members of the Stanford HCI Group, SALT Lab, and Situated Systems Reading Group for their helpful feedback. We also thank our study participants for their time and useful perspectives. This work was sponsored by the Stanford Institute for Human-Centered
Artificial Intelligence (HAI). Dora Zhao is supported in part by the Paul \& Daisy Soros Fellowship for New Americans. Michelle Lam
is supported by a Stanford Interdisciplinary Graduate Fellowship.
\end{acks}
\appendix
\bibliographystyle{ACM-Reference-Format}
\bibliography{egbib}
\appendix
\input{Sections/10_Appendix}
\end{document}

%% file: Figures/listing_style.tex
\colorlet{punct}{red!60!black}
\definecolor{background}{HTML}{FFFFFF}
\definecolor{delim}{RGB}{20,105,176}
\colorlet{numb}{magenta!60!black}

\lstdefinelanguage{json}{
    basicstyle=\footnotesize\ttfamily,
    stepnumber=1,
    numbersep=8pt,
    showstringspaces=false,
    breaklines=true,
    backgroundcolor=\color{background},
    literate=
     *{0}{{{\color{numb}0}}}{1}
      {1}{{{\color{numb}1}}}{1}
      {2}{{{\color{numb}2}}}{1}
      {3}{{{\color{numb}3}}}{1}
      {4}{{{\color{numb}4}}}{1}
      {5}{{{\color{numb}5}}}{1}
      {6}{{{\color{numb}6}}}{1}
      {7}{{{\color{numb}7}}}{1}
      {8}{{{\color{numb}8}}}{1}
      {9}{{{\color{numb}9}}}{1}
      {:}{{{\color{punct}{:}}}}{1}
      {,}{{{\color{punct}{,}}}}{1}
      {\{}{{{\color{delim}{\{}}}}{1}
      {\}}{{{\color{delim}{\}}}}}{1}
      {[}{{{\color{delim}{[}}}}{1}
      {]}{{{\color{delim}{]}}}}{1},
}

\lstdefinelanguage{Markdown}{
    basicstyle=\ttfamily\footnotesize, % specifies font size and line height
}

%% file: Sections/00_Abstract_v3.tex
A long-standing vision of computing is the personal AI system: one that understands us well enough to address our underlying needs. Today's AI focuses on what users do, ignoring \emph{why} they might be doing such things in the first place. As a result, AI systems default to optimizing or repeating existing behaviors (e.g., user has ChatGPT complete their homework) even when they run counter to users' needs (e.g., gaining subject expertise). Instead we require systems that can make connections across observations, synthesizing them into insights about the motivations underlying these behaviors (e.g., user's ongoing commitments make it difficult to prioritize learning despite expressed desire to do so). We introduce an architecture for building user understanding through \emph{behavior latticing}, connecting seemingly disparate behaviors, synthesizing them into insights, and repeating this process over long spans of interaction data. Doing so affords new capabilities, including being able to infer users' needs rather than just their tasks and connecting subtle patterns to produce conclusions that users themselves may not have previously realized. In an evaluation, we validate that behavior latticing produces accurate insights about the user with significantly greater interpretive depth compared to state-of-the-art approaches. To demonstrate the new interactive capabilities that behavior lattices afford, we instantiate a personal AI agent steered by user insights, finding that our agent is significantly better at addressing users' needs while still providing immediate utility.

%% file: Sections/01_Intro_v6.tex
Few ideas in computing have proven as persistently compelling, and as persistently elusive, as the personal AI system. Even over half a century ago, researchers in human-computer interaction were imagining computer agents that can assist users with their everyday tasks~\cite{kay1984computer,negroponte1970architecture}. This vision has expanded with proposals such as personal interface agents acting on behalf of their users~\cite{maes1994agents}, recommender systems filtering information to an individual's taste~\cite{resnick1994grouplens}, and adaptive interfaces reshaping in response to user actions~\cite{gajos2004supple}. 
Yet what continues to elude us are personal AI systems that can address our needs, without us explicitly spelling them out. 

Given the remarkable improvements in AI capabilities, why have we still not achieved these visions of personal AI? Today, systems focus on modeling \emph{observations} about the user, such as facts~\cite{shaikh2025creating,memories,karen2017contextual}, preferences~\cite{resnick1994grouplens,bai2022training}, or demonstrated actions~\cite{cypher1993watch,shaikh2025aligning,yang2025fingertip}.~\citet{kleinberg2024inversion} argue, however, that this approach is fundamentally insufficient: it suffers from an \emph{inversion problem} where AIs model our behavior but not our mental state. This limits the AI's focus to the ``what'', ignoring the ``why''--- hindering its ability to generalize or take appropriate proactive action. For example, while modern AI agents can complete concrete tasks (e.g., rescheduling a meeting), they are often so myopically focused on the task that they fail to address the underlying reason why we are doing that task in the first place (e.g., that we often overlook calendar conflicts and should have recognized the need to reschedule earlier). 

To achieve proactive personal AI, then, we require architectures that can act like a detective, synthesizing seemingly disconnected observations into underlying motives to explain someone's behavior. Rather than summarizing the facts of the user's behavior, these architectures must connect the dots to identify whether, for example, the user's behavior indicates anxiety around sending a draft to their collaborators, or whether the user is likely to forget about a social event---to guide the personal AI's actions to be most accommodating and helpful. To build user understanding like a detective would, one option could be to prompt language models with a trove of behavioral data, capitalizing on reasoning capabilities~\cite{guo2025deepseek,openai2024openaio1card}. However, models struggle to identify meaningful patterns when presented with large volumes of unstructured information, producing generic outputs or hallucinating~\cite{liu2024lost,lampinen2025latent}. Summarizing the data would reduce the precision of user understanding~\cite{chen2025compress,zhong2024memorybank}. Alternatively, prior work has retrieved recent ~\cite{park2023generative,ong2025towards} or similar behaviors~\cite{shaikh2025creating, rezazadehisolated,shi2025impersona}. But these strategies are akin to a detective only examining the last three events or evidence mentioning the same keyword. As a result, they end up grouping based on obvious characteristics, producing surface-level descriptions of the user.

We instead propose an architecture for building user understanding through what we term \emph{behavior latticing}. Behavior lattices produce user insights, inferences about the user's motivations, from observations. The lattice connects observations or lower-level insights to a set of higher-level insights in a many-to-many relationship, like a web. Those insights then get synthesized again to a higher level of the lattice, progressively producing more cross-cutting conclusions. We describe an algorithm that operates over rich, multi-day observations of user behavior to produce behavior lattices.\footnote{Project website and code implementation are available at \url{https://stanfordhci.github.io/lattice/}.} First, the algorithm organizes user observations so that each observation can belong to multiple groups and each group draws from multiple observations. For example, an observation that a user juggles administrative tasks suggests a tendency to overcommit when paired with volunteering for additional lab service, but ``productive procrastination'' when paired with observations of many unfinished high-priority tasks. Second, the algorithm ensures that behavior lattices are repeating. By repeating the structure hierarchically, behavior lattices can organize long periods of user data, thereby allowing us to contextualize which insights are recurring over time and which are tied to a particular setting.

Behavior latticing enables a new set of interactive capabilities for how personal AIs can understand users. First, by densely mapping observations across contexts and time, systems become able to act on the classic HCI truism of addressing users' underlying needs rather than narrowly solving their literal tasks~\cite{patnaik1999needfinding,rogers2023interaction,norman2013design}. Second, by linking granular observations, we can surface subtle patterns that accumulate over time, articulating aspects of user behavior that they themselves may not have been aware of. These two capabilities can further enable the design of personal AI systems across a breadth of application areas, including end-user customization of UIs~\cite{bolin2005automation} and social media feed curation~\cite{popowski2026social,malki2025bonsai}. 

To demonstrate this idea, we embed behavior lattices into a system \sysnamenospace, which steers a personal AI agent with user insights. We conduct a technical evaluation of our user insights, validating that our insights have interpretive depth while also being accurate. We collect \tratings ratings on outputs from our approach and from a state-of-the-art user modeling method~\cite{shaikh2025creating}. Our insights are rated as significantly deeper ($M = 1.17$ vs. $M=-0.37$ on a $-3$ to $3$ Likert scale) without sacrificing accuracy ($M = 1.30$ vs. $M=1.72$). Second, we recruited \pnum participants for an end-to-end evaluation, deploying \sysname to observe their computer usage for a minimum of 4 days. We synthesized user insights from this data and proposed actions \sysname could take on the user's behalf, collecting 140 ratings over 35 tasks. Our insight-steered actions were significantly better at addressing participants' underlying needs ($t=2.69$, $p=0.01$), while maintaining the same immediate utility.

In this work, we contribute an architecture for understanding the ``whys'' of user behavior through \emph{behavior latticing}. We describe a technical evaluation of our approach and a longitudinal evaluation of \sysnamenospace, a personal agent steered by user insights. Beyond agentic AI, this shift in perspective has implications for how we build personal applications more broadly, from end-user customization to content curation.

%% file: Sections/02_RelatedWork.tex
We build on existing work related to modeling user behavior and aligning AI systems using higher-level concepts (e.g., values). 

\subsection{Learning About Users}
\label{subsec:lit_learning}
The first challenge for building personal AI systems is gathering a requisite understanding of the user~\cite{fischer2001user,jameson2001modelling,horvitz1998lumiere}. One way to gather this information is by directly asking. The exact response elicited from users can take many forms, including a rating~\cite{bai2022training,li2024personalized}, natural language explanation~\cite{lieliciting,peng2025morae,vaithilingam2025semantic}, or demonstration~\cite{cypher1993watch,sugiura1998internet}. Prior works have demonstrated the efficacy of explicit preference elicitation particularly in high-uncertainty situations where user preferences cannot be reliably inferred~\cite{peng2025morae,hahn2024proactive,ma2025ambigchat}. Recent advances in long-context modeling have further enabled systems to retain and condition on large amounts of user-provided information across extended interactions~\cite{gao2025train,warner2025smarter}. These methods assume that users are able to express all of the requisite information to the model. However, moving from more objective observations (e.g., facts, preferences) to the higher-level concepts, that form our user insights, leads to gaps in articulation~\cite{patnaik1999needfinding}.

Another option is to learn these preferences through users' interaction with the system. This approach is akin to how existing chatbot services develop an understanding of the user (e.g., ChatGPT's ``Memory''~\cite{memories}) or infer user intent from underspecified queries~\cite{berant2025learning,kim2026discoverllm,kim2024beyond,choi2025bloomintent}. Other methods induce graphical summaries of users' interactions with applications, such as digital creativity support tools, to better understand their activity~\cite{smith2025fuzzy,lee2024impact,goldschmidt2014linkography}. Beyond individual-level signals, methods, such as collaborative filtering~\cite{resnick1994grouplens} allow us to learn user representations based on patterns in interactions between other users and items that the system has already seen. While this class of methods does not require the user to explicitly specify information about themselves, it does constrain the context we can learn from to a narrow window of interaction. 

Finally, a growing body of work has advocated for learning about users through passive observation across many contexts. For example, \citet{shaikh2025creating} introduce a method for learning a ``General User Model'' (GUM) that captures users' behaviors and preferences by processing their computer interaction data. Other works have introduced similar approaches across a suite of different inputs, including audio recordings, mobile device interactions, and wearable data~\cite{yang2025fingertip,yang2025contextagent,danry2026mindmap,arakawa2024prism,pu2025promemassist}. 

The contribution of behavior latticing is not only what we learn about users but how we do so. Across user modeling methods, we find a common two-step paradigm: identify a group of related behaviors, then interpret~\cite{shaikh2025creating,park2023generative,danry2026mindmap,zhong2024memorybank,zulfikar2024memoro}. Behavior latticing differs at both steps. First, when grouping, most methods compress observations whether through only retrieving ``relevant'' behaviors --- where relevance can be defined by similarity~\cite{shaikh2025creating,shi2025impersona,rezazadehisolated,zulfikar2024memoro}, importance~\cite{park2023generative}, or recency~\cite{park2023generative,ong2025towards} --- or summarizing inputs into a condensed format~\cite{lam2024concept,chen2025compress,zhong2024memorybank}. Instead we group over all observations within a temporal sequence (e.g., collected over one day), forming overlapping connections across behaviors. This design not only affords access to more context but also allows more flexibility in what gets grouped. As a result, our architecture surfaces groupings existing methods structurally miss: contextually unrelated observations demonstrating the same motive, individually unimportant actions that show a deeper pattern, or groups that span across time horizons. Next, in the interpretation stage, existing methods either operate at a single level of abstraction --- refining conclusions based on new low-level observations~\cite{shaikh2025creating,danry2026mindmap} --- or, when they do maintain hierarchy, build it through local operations like routing to similar nodes~\cite{rezazadehisolated} or mixing observations and inferences in a shared pool~\cite{park2023generative}. In both cases, deeper interpretation is not guaranteed. Instead, latticing enforces higher-level inferences \emph{by design}. Each layer operates only on the outputs of the layer below, progressively deepening interpretation.

\subsection{Going Beyond Observed Behavior}
Across a wide range of personal AI applications, existing work has wrestled with the gap between the behaviors users reveal and the outputs that would actually benefit them~\cite{milli2021optimizing,cheng2026sycophantic,khambatta2023tailoring}. For example, in recommender systems, optimizing for behavioral signals, such as clicks, dwell time, or likes,  remains the de facto practice; however, research has shown doing so can significantly lower user utility~\cite{besbes2024fault,milli2025engagement}. Several approaches have sought to address this, whether by surfacing niche content that users would not discover on their own~\cite{besbes2024fault}, directly incorporating signals for user enrichment over engagement~\cite{anwar2025recommendation}, or aligning content with higher-order constructs such as human values~\cite{jahanbakhsh2025value, kolluri2025alexandria}. Similar ideas have emerged for shaping large language models, such as trying to align them with broader human values rather than only considering user preference~\cite{bai2022constitutional,hendryck2021saligning,sorensen2024value,shen-etal-2025-valuecompass,ellis2025training}. In more domain-specific applications, systems are designed around pre-specified principles, such as tutoring applications that adhere to pedagogical best practices rather than simply responding to students' observed performance~\cite{team2024learnlm,scarlatos2025training}. These works largely rely on the designer or researcher to prescribe what higher-level objectives the system ought to align to, be it a construct like ``user empowerment''~\cite{ellis2025training} or a set of pedagogical principles~\cite{scarlatos2025training}. However, which values are most apt or which principles applicable depends on the user. Rather than relying on fixed objectives, our work presents an architecture that produces insights specific to each user, bridging surface-level behavioral data and the underlying motivations that ought to inform personal AI systems.

%% file: Sections/03_Architecture_v1.tex
\begin{figure*}[htbp!]
    \centering
    \includegraphics[width=\linewidth]{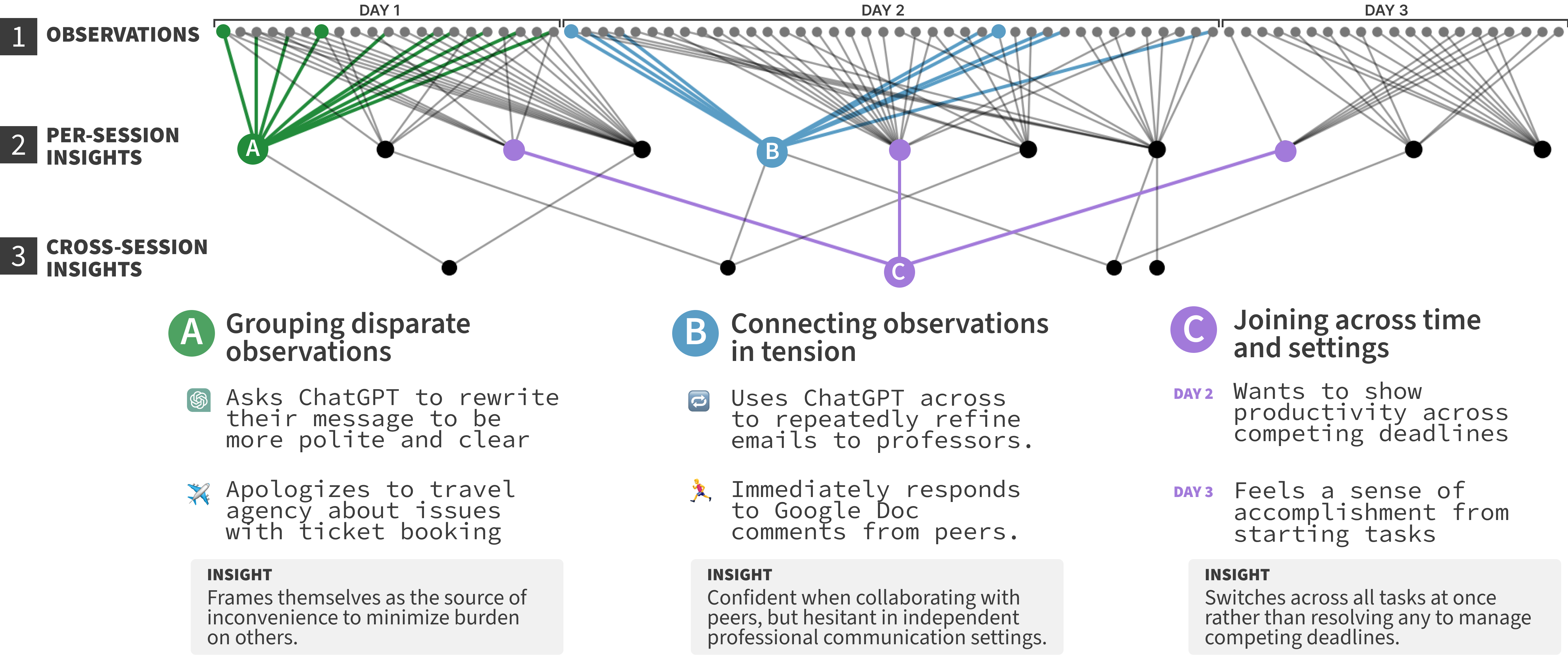}
    \caption{Modeling user behavior as a \emph{lattice} enables the following capabilities: (A)~connecting observations from different applications or contexts that share a latent motivation, (B)~juxtaposing observations that are in tension, and (C)~linking patterns that recur across time. We illustrate these capabilities using actual data from a participant in our technical evaluation.}
    \label{fig:latticing}
\end{figure*}
Should users specify a complete biography of themselves to their AIs in order to achieve personalization benefits? This approach requires users expend significant articulatory effort to first reason about and then verbalize insights for each interaction. In addition, and more critically, users often struggle with articulating their needs or motivations even when asked~\cite{patnaik1999needfinding,leonard1997spark,christensen2005marketing,popowski2026social,pommeranz2010user,katz1969introduction}. Given this, having a method for learning insights becomes especially important as the outputs that we target --- unlike factual descriptors --- are difficult to surface through direct elicitation. In this section, we first define our goal of \emph{user insights} before introducing our architecture for creating these insights via latticing.

\subsection{Desiderata for User Insights}
User insights can be a broad term in HCI, so we define a more focused version \textbf{user insight} for our purposes: an inference about the user's underlying motivations. There are two criteria to which user insights must adhere:
\begin{enumerate}[leftmargin=*]
    \item \textbf{Accuracy}: Insights about the user should be true. Inaccurate insights risk leading to AIs that are unuseful, or worse, actively harmful. While accuracy is necessary, it is not sufficient. For example, the statement ``Amy uses Overleaf to write her paper'' is accurate but does not constitute an insight as it fails to provide an inference about her motivation. 
    \item \textbf{Depth}: The second criterion for a user insight is depth. Drawing on the criteria commonly laid out for user insights in design research, we define ``depth'' as providing explanation or motivation for user behavior rather than simply stating the surface-level observations of what the user is doing~\cite{patnaik1999needfinding,paul2023insight}. For instance, the statement ``Amy prefers Overleaf because it reinforces her identity as a `serious' academic'' offers a more in-depth interpretation of her motivation.
\end{enumerate}

\noindent Achieving both is challenging. Existing approaches prioritize accuracy, reporting objectively true or false information that can lack depth. However, as we demonstrate, pursuing only accuracy without depth leads to limited personal AIs that only focus on what the user is doing. Conversely, generated outputs can have depth but lack accuracy, hallucinating plausible but unsupported conclusions. 
\subsection{Producing User Insights}
\label{subsec:insight_pipeline}
Creating user insights that encompass both accuracy and depth requires more than just summarization: it requires synthesizing seemingly disparate observations over long time periods into underlying explanatory hypotheses. To achieve this goal, we describe an architecture for inferring user motivations from unstructured interaction data via \emph{behavior latticing}. The term \emph{lattice} refers to two key properties of our architecture: (1)~user behaviors are connected in an interweaving fashion and repeatedly across sessions, and (2)~the insights resulting from one layer of the lattice are then grouped and used as input to further layers of the lattice, producing more cross-cutting interpretations that form vertical layers.

Starting with a set of unstructured data about the user (e.g., chat logs, audio recordings), this approach outputs user insights through repeatedly connecting and interpreting observations of user behavior (Fig.~\ref{fig:insight_pipeline}). Our architecture is agnostic to the modality of input data so long as it is capturing rich user behavior over an extended period of time. In this section, we demonstrate our pipeline with a simple example of inputting a user's ChatGPT conversation history. In subsequent sections, we expand the scope of input data to large-scale, longitudinal screenshots of the user's computer usage. 

\subsubsection{Making Observations}
\label{subsubsec:observations}
The first step of our architecture requires \textbf{observations} about the user. Observations are factual descriptions of user behavior, such as what actions they are doing, who they interact with, or which tools they use. We begin by processing a session of input data, which we define as a bounded unit of user interaction. For instance, since the input data in the example are chat logs, session-level observations come from messages in a single chat thread, and cross-session insights are derived from all conversations between the user and the chatbot. 
\begin{quoting}[leftmargin=16pt, rightmargin=8pt]
\small
Here is part of the conversation: \\
\texttt{
Amy: be honest. how is this discussion for a paper. am i capturing enough nuance?\\
ChatGPT: This discussion could benefit from a few changes. Here's a new draft.\\
Amy: actually let's stick to refining. this is too pretentious.}
\end{quoting}

\noindent Our goal is to understand what the user is doing in the moment. To this end, we employ language models to form discrete observations about the data (e.g., actions~\cite{yang2025fingertip,lyu2026personalalign,yang2025contextagent}, affect~\cite{nasoz2007affective,martinho1999cognitive}, preferences~\cite{shaikh2025creating,singh2025fspo}). 

\begin{quoting}[leftmargin=16pt, rightmargin=8pt]
\small
\noindent
We provide an example of an observation.\\
\texttt{
Amy requested a rewrite to be `more CSCW-ready and theoretically grounded,' but then corrected `let's stick to refining. this is too pretentious.'}
\end{quoting}

\subsubsection{Forming Insights via Behavior Latticing}
\label{subsec:latticing}
Interpreting user behavior requires making connections across ostensibly disparate and individually inconsequential observations. For example, a user toggling between three different note-taking apps, spending an hour configuring keyboard shortcuts, and abandoning a half-written outline to start a new one are unrelated at face value. However, together they paint a picture of someone who is more energized by setting up productivity systems than by using them. 

We construct the \emph{behavior lattice} in a bottom-up approach. To start, we use observations as inputs, forming the leaf nodes of the behavior lattice (Fig.~\ref{fig:latticing}~\blacksquarelabel{1}). Using a reasoning model, we synthesize observations, producing the set of user insights in layer \blacksquarelabel{2}. We apply this process recursively, synthesizing insights to make new layers (e.g., insights in \blacksquarelabel{2} are the input for \blacksquarelabel{3}).

We formalize our method using the following notation. A behavior lattice consists of $n$ layers $\mathcal{L}_0, \mathcal{L}_1, \ldots, \mathcal{L}_{n-1}$. The base layer, $\mathcal{L}_0$ refers to the set of observations forming the lattice's leaf nodes. Subsequent layers $\mathcal{L}_j = \{\ell_{j,1}, \ell_{j,2}, \ldots, \ell_{j,k}\}$ are the sets of user insights produced by a reasoning model. Each insight $\ell_{j,k}$ consists of a title, a brief description, and the set $S_{j,k}$ of evidence from the layer below that supports the inference in $\ell_{j, k}$. The edges of our lattice connect an insight $\ell_{j,k}$ to each piece of evidence in $S_{j,k}$.

\noindentparagraph{Adding a new layer to the lattice.}
To add a new layer, we identify sets of supporting evidence within the latest layer, $\mathcal{L}_{n-1}$ and synthesize them into insights, forming $\mathcal{L}_{n}$. We prompt the reasoning model to group together elements in the layer ($\ell_{n-1, 1}$, \ldots, $\ell_{n-1, k}$) that are in tension, contradictory, or represent a recurring pattern. These groups are the supporting evidence sets $\{S_{n, 1}, \ldots, S_{n, k}\}$. Given $S_{n,k}$, synthesis follows naturally by leveraging the reasoning capabilities of models. Our prompt guides the model to draw connections within $S_{n,k}$ and infer user motivation, which is stored as $\ell_{n, k}$. 

\noindentparagraph{Many-to-many mapping within a layer.} An important property of the lattice is the dense set of connections between layers. We do not constrain observations to map only to a single insight, or lower-level insights to map only into one higher-level insight. For example, if we consider the edges between observations in $\mathcal{L}_0$ and insights in $\mathcal{L}_1$, the model can assign the same observation as evidence for multiple insights, if relevant. The number of insights that an observation maps to emerges from the model's reasoning rather than from a predefined constraint. As a result, some observations are linked to almost all insights, indicating a particularly unique or informative behavior. Conversely, there are observations not linked to any insight, as we do not enforce an exhaustive mapping.

\begin{quoting}[leftmargin=16pt, rightmargin=8pt]
\small
\noindent
    Our architecture links the following observations: \\
    \texttt{[1. Amy requested a rewrite to be more CSCW-ready and theoretically grounded.}, ..., \texttt{5. She repeatedly fine-tunes ChatGPT's outputs to match her own tone.]} \\
    From these observations, we infer that Amy wants to speed up her writing process with AI, but this ends up conflicting with her strong personal voice
\end{quoting}

\noindent
Our current implementation only builds the lattice upward, going from observations to form user insights. However, given that this architecture produces a graph, other traversal algorithms are applicable. Advancements on our architecture could combine upward passes (as we do now) with downward passes. Propagating generated insights downward could allow the system to reexamine how observations are grouped, or even what observations are made.

\subsubsection{Applying Latticing Recursively}
\label{subsec:recursive}
A lattice with two layers (i.e., input observations and one layer of insights) still produces a set of user insights, but it lacks information on the generalizability of the insights in the user's life. For example, insights produced from Amy's chat history about her paper may only apply to academic writing. Or, we might see this similar patterns related to learning new materials or creating presentations. Right now, we lack sufficient context to draw any conclusions. 

To address this challenge, our approach builds additional layers to the lattice, drawing connections across increasingly longer time horizons. For example, we can link insights that indicate the same motivation but occur across temporally distant sessions, or identify insights that may be in tension with each other. Through this step, we can contextualize which insights are more enduring (i.e., occur across multiple sessions and settings) versus those that are more context specific (i.e., arising in a single session or setting). 

After applying behavior latticing, we use the insights in the final layer, $\mathcal{L}_{n-1}$, as our output. In addition to the textual description of the inferred motivation, the insight also includes the contexts it applies to (e.g., the insight may only be related to professional, not personal interactions), and supporting evidence comprising all of the descendents in the lattice, from intermediary insights to observations.

\begin{quoting}[leftmargin=16pt, rightmargin=8pt]
\small
\noindent
\texttt{\textbf{Insight}: Amy balances a desire to demonstrate theoretical sophistication with concerns about being perceived as inaccessible.\\
\textbf{Context This Applies}: Presenting academic work\\
\textbf{Support}: [Connected insights and observations]}
\end{quoting}

\subsubsection{Architecture Parameters} \new{Behavior latticing is parameterized by session length and number of lattice layers. We define a session as a single chat thread for conversational data and a calendar day for computer usage data, representing natural units of interaction for each data type. We set the number of layers to 3 (i.e., observation layer, per-day insights, and insights across days). While multiple layers deepen interpretation (Sec.~\ref{subsec:recursive}), too many risk producing overly abstract insights. Ultimately, the parameters depend on the richness of the input data and the downstream application.}

%% file: Sections/04_Dawn.tex
\begin{figure}[t!]
    \centering
    \includegraphics[width=0.75\linewidth]{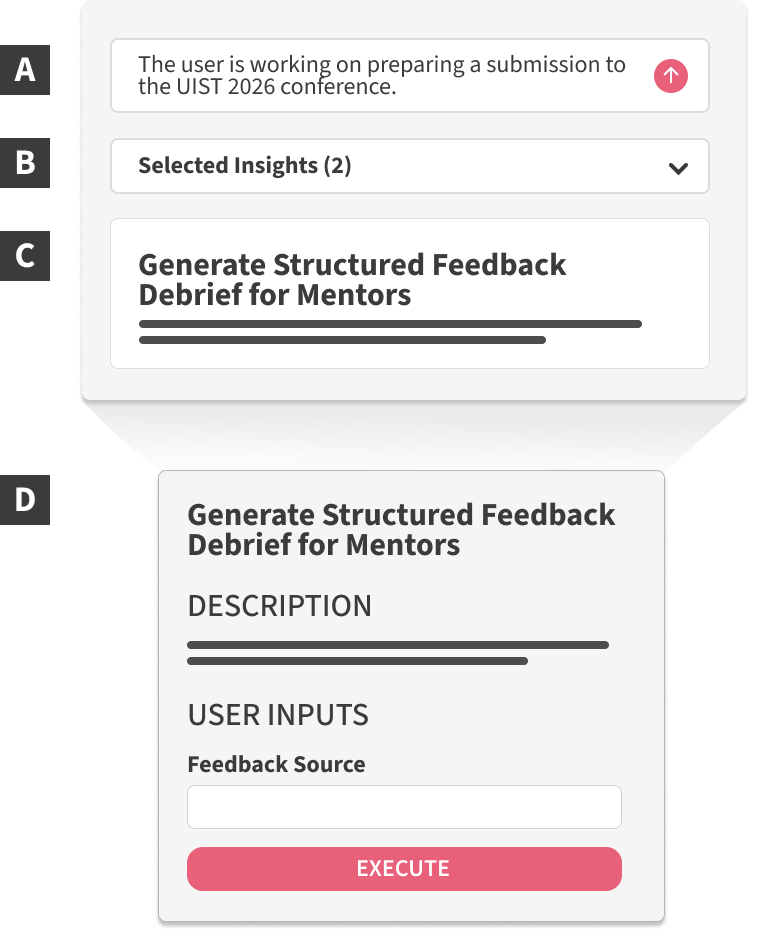}
    \caption{\sysname is an AI agent that discovers tasks where the user requires personal assistance (A). We use insights from the behavior lattice (B) to propose actions that the agent can take (C). The user can provide additional information before deploying the agent (D).}
    \label{fig:dawn}
\end{figure}

Next, we demonstrate how to embed behavior latticing into one likely application area: personal AI agents. Concretely, we present our system \sysname (Fig.~\ref{fig:dawn}),\footnote{We name our system \sysname to evoke both looking toward the longer horizon in assisting users and the moment of realization (``it dawned on me'') that participants reported when seeing generated insights.} a personal agent steered via user insights. \sysname first produces user insights from observations of users' computer use. Then, it proactively induces tasks in which the user requires the assistance of a personal agent and proposes actions that are informed by these insights. 

\subsection{System Components}
\sysname consists of three core components. First, we deploy behavior latticing to produce insights about the user from screenshots of their computer usage. Second, \sysname uses insights to guide what actions the agent should take to address the user's needs. Finally, a tool-calling MCP agent executes the actions.

\subsubsection{Generating Insights from Screenshots}
To form a rich understanding of the user, we capture screenshots of their computer usage, which is input to our insight generation pipeline (Sec.~\ref{subsec:insight_pipeline}). Our current implementation leverages the \texttt{Screen Observer} from \citet{shaikh2025creating}, taking screenshots of the user's screen based on input monitoring (e.g., keystrokes, mouse clicks). We use a VLM to extract a transcription of the user's screen and summary of their actions from the screenshots.

\subsubsection{Proposing Insight-Guided Actions}
We select tasks where having a personal agent offers utility over a generic agent. Critically, this distinction is user-dependent. Writing a sorting function may not require personal assistance for a software engineer, but might for someone who is just learning to program. To implement this, we start by inferring the low-level tasks the user is working on also from screenshots. After compiling a list of these low-level tasks, we then instruct a reasoning model to synthesize a set of higher-level tasks that represent what the user is working on for that day. We filter tasks using an LLM classifier, scoring each one from 0 to 1 based on the estimated utility personal assistance would provide relative to a generic agent response, conditioned on the insights we already have about the user~\cite{horvitz1999principles}. Only tasks exceeding a threshold of 0.75 are retained.\footnote{We select 0.75 based on pilot testing, prioritizing precision in identifying which tasks require personal assistance.}

For each task, we retrieve the top two most relevant insights using a cross encoder (Voyage AI’s \texttt{rerank-2.5})~\cite{voyage2024rerank}. \new{In line with prior work, we found including too many insights led to generic outputs, likely because the model is trying to optimize across many competing objectives~\cite{lam2025just}.} With the insights and task at hand, we then prompt an LLM to propose \emph{actions} that a tool-calling AI agent could execute. To ensure actions are feasible, we provide a set of implementation constraints to the LLM specifying criteria to which the solutions must adhere. 

\subsubsection{Executing Agent Actions}
Finally, \sysname can execute proposed actions through deploying ReAct agents with tool-calling abilities~\cite{yao2022react}. \sysname has access to read/write access to the user's Google Drive, local filesystem, and Apple Calendar, as well as web search capabilities, selected to cover a representative set of tools that a typical ``personal assistant'' might have. 

\subsection{System Implementation}
\sysname is deployed as an Electron desktop app using React. All images are first processed using a local OCR model\footnote{https://github.com/JaidedAI/EasyOCR}; those that contain information from sensitive URLs are not saved. We use four different LLMs in our system, selected for their performance on the task and for data security. We use Gemini 2.0 Flash Lite to transcribe and summarize screenshots, Claude Sonnet 4.6 for generating user insights, and Claude Sonnet 4.5 for proposing agent actions. Finally, we deploy agents using Gemini 3 Pro with the DSPy framework~\cite{khattab2023dspy}. All models are accessed via institution-provisioned servers. See Appendix for more details.

%% file: Sections/05_Eval.tex
We conduct two evaluations. First, recalling our original design goals of \emph{depth} and \emph{accuracy}, our technical evaluation validates that our insights provide more interpretative depth compared to existing methods while still being accurate (Sec.~\ref{sec:technical_eval}). Second, in Sec.~\ref{sec:e2e_eval}, we conduct an end-to-end evaluation of \sysname to see whether insight-steered actions better address underlying needs. All studies were approved by our organization’s IRB.

\subsection{Procedure} We recruited participants via mailing lists and word-of-mouth. For both evaluations, participants downloaded \sysname and recorded their screen activity for a minimum of four days. 

\subsubsection{Technical Evaluation}
We compare insights (i.e., \textbf{\textsc{Insights}}) produced by our behavior latticing architecture to outputs from existing user modeling methods (i.e., \textbf{\textsc{Observations}}). We use statements about users produced from \citet{shaikh2025creating}'s General User Models (GUM) as our exemplar of an Observations-focused system. \new{We select GUM as it is one of the few methods that construct user models from rich unstructured interaction data, affording the ability to observe the user across a breadth of contexts. As such, GUM represents a leading approach in observational user modeling, making it a natural point of comparison for evaluating what new capabilities behavior latticing enables.}

We recruited \tnum participants (P1-P9) for this evaluation. Each participant rated an equal number of \techours{} and \techbase{} produced from their screen activity. Statements 
were presented in two counterbalanced blocks with block order randomized across participants. In total, we collected \tratings ratings.

\subsubsection{End-to-End Evaluation} We recruited \pnum participants (P10-P21), who are distinct from those who participated in our technical evaluation. For the evaluation, participants rate proposed agent actions. They also complete an optional 30-45 minute interview during which they rated user insights and executed agent actions, in addition to providing holistic reflections. Interviews are recorded and transcribed with Dovetail.\footnote{https://dovetail.com/} We interviewed all but one participant. Participants were compensated with a \$50 Tremendous gift card with an additional \$25 for the interview. 

We used the first three days of screen observations as input to our insight generation pipeline; we used the last day to identify tasks that our agents can assist users with. Participants were asked to rate proposed agent actions across three tasks.\footnote{If participants had fewer than three tasks that were classified as requiring a personal agent, then all available tasks above the threshold were used} Examples of tasks include ``writing an academic research paper,'' ``improving competitive programming skills,'' and ``providing constructive feedback for a group project evaluation.'' See Appendix for full list.

For each of these tasks, we generated actions for two conditions. First, our system, \textbf{\sysnamenospace}, is steered by relevant user insights. Second, the \textbf{\textsc{Baseline}} agent is steered using relevant context about the task (e.g., if the task is ``writing a grant proposal'', relevant context would be ``user is co-writing the proposal with her advisor'' and ``the proposal is about quantum physics''). \new{We design \eebase{} to reflect the standard configuration of popular personal agents, such as OpenClaw~\cite{steinberger2026openclaw}, which default to using immediate, task-specific context to guide execution.} In total, we collected ratings on 140 actions (70 per condition) across 35 tasks.  

\subsection{Measures}
We assess the quality of the statements produced about the user (\textsc{Insights} and \textsc{Observations}) and their impact on agent actions. 

\subsubsection{User Insights}
Participants rated generated statements, \textsc{Insights} and \textsc{Observations}, on two dimensions using a 7-point Likert scale ($-3$: Strongly Disagree to $+3$: Strongly Agree). The dimensions are \textit{accuracy} (``this statement about me is accurate'') and \textit{depth} (``this statement reveals something important about who I am, not just what I do''). Participants were also invited to provide qualitative reflections on statements. 

\subsubsection{Proposed Actions}
First, participants rate how important the task is to them on a 7-point Likert scale from $-3$: Very unimportant to $+3$: Very important. Then, for each proposed agent action, participants provided the following ratings on a 7-point Likert scale ($1$: Not at all to $7$: Extremely). Participants rate the (1)~\emph{immediate utility} of the action for assisting with the task as well as the extent to which the action addresses their (2)~\emph{underlying personal needs} related to the task. As an exploratory measure, we also ask participants to rate the (3)~\emph{novelty} of the action, an index defined as the average of two questions: the extent to which the action would change their strategy for the task, and how likely they would have been to think of the action on their own (Cronbach's $\alpha=0.87$).

\subsubsection{Executed Actions}
For each participant, the agent executes one \sysname action and one Baseline action for a randomly sampled task. Participants rated \emph{completion} on a binary scale (i.e., did the agent execute what was described) and \emph{quality} of execution on a 7-point Likert ranging from $-3$: Very Poorly to $+3$: Very Well. 

%% file: Sections/06_TechnicalEval.tex
\input{Figures/ComparingInsights}
In this section, we report ratings on accuracy and depth for \textsc{Insights} versus \textsc{Observations}. We also include reflections on \textsc{Insights} generated for participants in the end-to-end evaluation.  

\begin{figure}
    \centering
    \includegraphics[width=\linewidth]{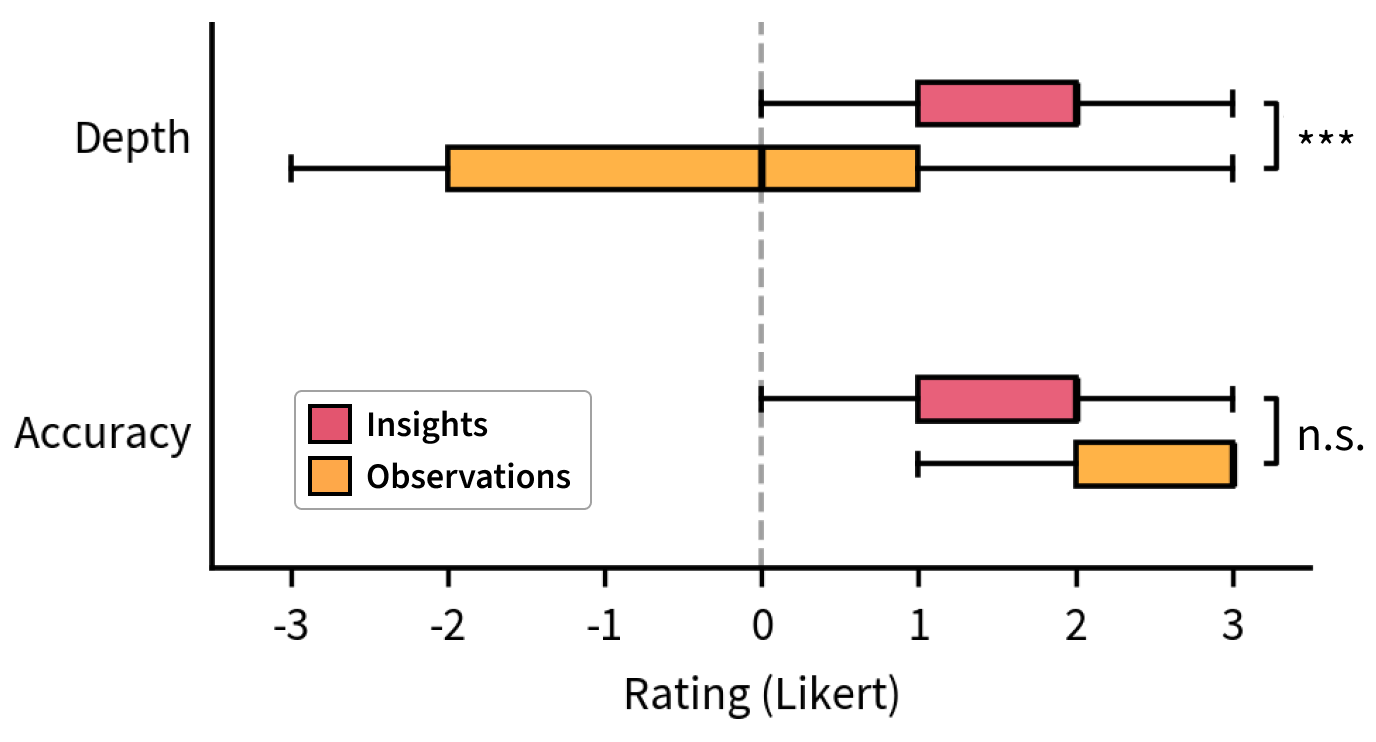}
    \caption{\techours{} are significantly deeper compared to \techbase{} ($t=7.78$, $p < .001$), meaning participants agree that the statements reveal something important about their identity, without compromising accuracy ($t=-1.76$, $p =0.08$).}
    \label{fig:technical}
\end{figure}

\subsection{Results}
We compare ratings of depth and accuracy for \techours{} and \techbase{}. In addition, we analyze the types of inferences made in \techours{}. Finally, we note limitations of our current architecture.

\noindentparagraph{\techours{} provide more in-depth understanding of users.} 
Compared to \techbase{}, \techours{} are deeper with a mean rating of $1.17$ versus $-0.37$ (see Fig.~\ref{fig:technical}). In other words, participants ``somewhat agreed'' or ``agreed'' that \techours{} were deep, but were neutral or ``somewhat disagreed'' regarding \techbase{}. A paired t-test confirms the statistical significance of this difference  ($t = 7.78, p < 0.001, d=0.95$).\footnote{Analyses using non-parametric statistical tests and mixed-effect models which yield qualitatively similar results (see Appendix).} Overall, participants rate \techours{} as deep $75.9\%$ of the time (rating $\geq 1$). When asked to compare outputs of both approaches, multiple participants (P1, P2, P3, P5) characterized \techbase{} as reporting ``\emph{factual knowledge''}, whereas \techours{} captured something more personal --- their ``\emph{personality''} (P1, P2) or \emph{``learned traits''} (P1). See Table~\ref{tab:gumvours} for examples.

We further examine what types of insights are generated through automated thematic  clustering~\cite{tamkin2024clio,lam2024concept}. Examples of emergent themes include how factors like social comparison influence participants' motivations (e.g., ``P21 is motivated by observing peers' progress''), tensions between what participants plan versus how they execute (``P14 aspires to deeply engage with academic material but relies on AI assistance when time limited''), and patterns about how they manage their workload (``P8 relies on meticulous tracking systems which can increases her workload and complexity under time pressure''). Full results in the Appendix.

\noindentparagraph{\techours{} provide deeper understanding without sacrificing accuracy.} 
Given that we apply a layer of interpretation rather than reporting objective observations, we expect our approach will be less accurate---it takes bigger risks in applying an interpretative lens onto the facts. \techours{} are rated as less accurate ($M=1.30$ vs. $M=1.72$). While there is a small effect ($d=-0.23$), the difference is not statistically significant ($t=-1.76$, $p=0.08$). On average, participants still ``somewhat agree'' or ``agree'' that \techours{} are accurate. Examples of accurate insights include inferences that P21 pursues multiple commitments without a single clear focus and P2 exists in a constant state of preparation as a way of deferring intellectually challenging tasks. In contrast, inaccurate \techours{} stemmed from incorrect interpretations, such as overindexing on P3's unconscious behaviors leading to the conclusion that she engages in tangential activities like browsing social media or booking personal appointments when she feels pressure at work. 

\noindentparagraph{\techours{} surface information that users were unaware of.} 
We start with the following quote from P12:
\begin{quoting}[leftmargin=20pt, rightmargin=20pt]
\small
    ``[This \textsc{Insight}] talks about a thing that I don't think I've put into words quite as well as it did but have regularly talked to my spouse about how this affects my life...I think this is something that would take a therapist multiple meetings with me to reach this conclusion.''
\end{quoting}
\noindent 
Other participants, such as P21 had similar experiences: ``\emph{The way I view myself and perceive myself is more blurry, but [seeing the insights] is like looking at a mirror and finally understanding.}'' \techours{} resonated with users beyond the study. Participants mentioned sharing them with friends, sparking conversations about whether and how they exhibited these patterns in daily life. Several requested copies to keep for themselves. As P12 quipped, ``\emph{I should just screenshot this and send it to my husband and be like, are you aware of all of these insights about me?}''

Participants' reflections help explain \emph{why} \techours{} feel novel. For P20, he noted \techours{} drew upon unconscious behaviors, such as ``\emph{what your tone is, how much effort or how much attention you apply.'}'. Grouping these together led to a statement about himself that he had not considered before. P13 pointed out an \textsc{Insight} that linked checking Instagram likes, competitive coding rankings, and peer salaries --- behaviors he was aware of but would not have connected. This led to an inference that his self-assessment is often based on external metrics leading to frequent comparisons. In both cases, \emph{how} behaviors are connected can lead to \techours{} that are resonant and novel.

\subsection{Errors and Boundaries}
We identified two limitations: (1)~incorrect normative judgments about observed behavior and (2)~idiosyncratic conclusions arising from the limited time window.

\noindentparagraph{\techours{} can draw incorrect normative conclusions.}
Polarizing examples arise when the system makes incorrect normative claims. For example, P10 mentioned that one \textsc{Insight} placed the time he spent cooking with friends in tension with his research, making the normative claim that this tension is detrimental. This prompted reflection: ``\emph{I was like, do I have a problem?...Is my life in disarray because I cook too much?}'' He agreed that these two activities are in tension but disagreed with the normative conclusion. Adding directionality is essential for building systems that do not simply reinforce what the user is already doing. However, the risk of paternalism makes the importance of user input and feedback salient.

\noindentparagraph{What we learn is limited by the time window of observation.} Other inaccuracies stem from the limited time window of observation. For example, P14 noted several \techours{} were skewed since the system only observed her during spring break. The statements reflected \emph{``who [she] was at that moment}'' but was incongruent with how she views herself. Our approach can support longer timescales of observation data, and these results suggest that longitudinal data could be useful to counter spurious or time-bound inferences.

%% file: Figures/ComparingInsights.tex
\begin{table}[]
    \centering
    \footnotesize
    \begin{tabular}{>{\RaggedRight\arraybackslash}p{4cm}
    >{\RaggedRight\arraybackslash}p{4cm}}
    \toprule
    \textbf{\techbase{}} & \textbf{\techours{} (Ours)}\\
    \midrule
    P9 is an active PhD applicant for the 2025 cycle. & P9 demonstrates strong agency when helping others or collaborating, but experiences avoidance when facing decisions about her own career trajectory.\\
    P8 uses tomatotimers.com as part of his work session routine. & P8's productivity tools signal an intent to work but do not always direct behavior. They are often reactive measures used after many tasks have accumulated, rather than proactive systems.\\
    \bottomrule
    \end{tabular}
    \caption{\techours{} produced from our architecture present qualitatively different results than \techbase{} from existing user modeling methods~\cite{shaikh2025creating}.}
    \label{tab:gumvours}
\end{table}

%% file: Figures/ComparingActions.tex
\begin{table*}[t]
\centering
\footnotesize
\begin{tabular}{
    >{\RaggedRight\arraybackslash}p{2.2cm}
    >{\RaggedRight\arraybackslash}p{4.2cm}
    >{\RaggedRight\arraybackslash}p{4.9cm}
    >{\RaggedRight\arraybackslash}p{4.8cm}}
\toprule
\textbf{Task} & \textbf{Relevant Insights} & \textbf{\sysname Action} & \textbf{\eebase{} Action} \\
\midrule
Writing a comprehensive project report on AI and social media. 
&
1. Uses AI tools and external resources not to increase speed but to address uncertainty in her work\newline
2. Rapid platform-switching and social media browsing serve to manage difficult tasks, though this behavior can become more frequent when work feels uncertain
&
\textbf{Focused Work Session Structurer:} The agent will break down the project report work into discrete  micro-tasks with completion criteria and scheduled breaks. This reduces ambiguity about what to do next and creates natural stopping points for cognitive breaks, preventing unproductive task-switching.
&
\textbf{Report Outline with Source Integration:} The agent will query an LLM to create a comprehensive academic report outline on AI and social media, including section breakdowns, key points to cover for each complex concept, and suggested areas for discussing societal impacts.
\\
\midrule

Creating a Ph.D. research presentation.
&
1. Uses AI not only for productivity but as a way to manage the pressure of addressing knowledge gaps and accessing simulated mentorship.\newline
2. Demonstrates persistence and methodical creativity in structured competitions but seeks external guidance in ambiguous social or bureaucratic contexts.
&
\textbf{Generate Simulated Committee Q\&A Session:} The agent will analyze the user's presentation materials and research context to generate a comprehensive list of potential questions her PhD committee might ask, organized by difficulty level and topic area, along with suggested response frameworks for each question.
&
\textbf{Generate Research Presentation Structure:} The agent will analyze the user's research documents and create a comprehensive presentation outline with suggested content for each section, including introduction, methodology, findings, conclusions, and implications. The agent will structure the content appropriately for a doctoral committee review.
\\
\bottomrule

\end{tabular}
\caption{We provide examples of actions from \sysname versus \eebase{} for the same task and the insights used to steer \sysnamenospace.}
\label{tab:insight_actions}
\end{table*}

%% file: Sections/07_ActionEval.tex
\label{sec:end2end_eval}
Next, we examine how \sysnamenospace's actions compare to actions generated with relevant task context (\eebase{}). We also evaluate how well \sysname can execute proposed actions.

\begin{figure}
    \centering
    \includegraphics[width=\linewidth]{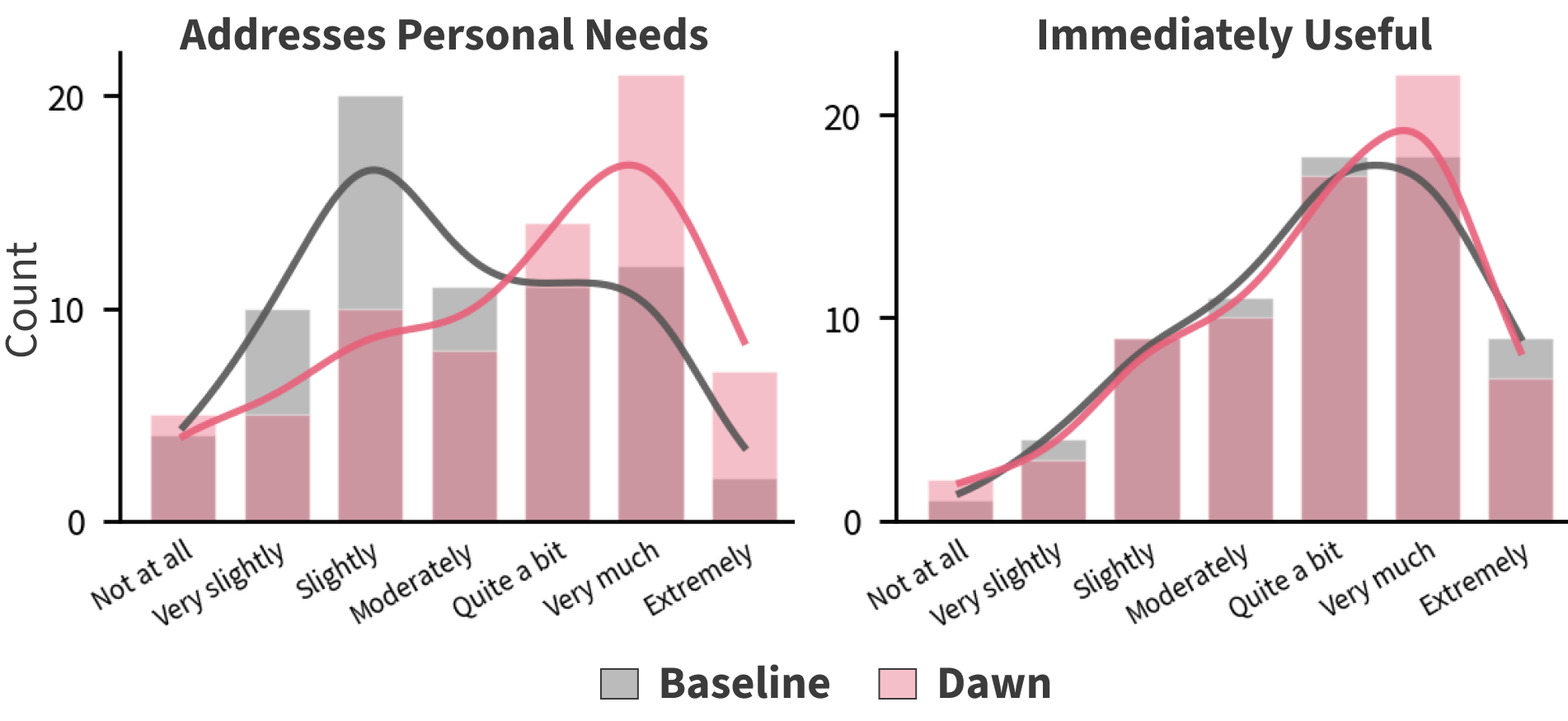}
    \caption{\sysname is better at addressing participants' underlying needs ($t=2.69, p = 0.01$) while retaining similar levels of utility ($t=0.0, p=1.00$). We plot the distributions of ratings across 140 actions, with smoothed density overlays.}
    \label{fig:e2e_results}
\end{figure}

\noindentparagraph{\sysname better addresses underlying needs, without sacrificing utility.} Participants reported that \sysnamenospace's actions are significantly better at addressing their underlying needs ($4.60\pm1.72$), compared to \eebase{} actions ($3.84\pm1.56$) (see Fig.~\ref{fig:e2e_results}). \sysnamenospace's actions are rated as addressing needs ``Moderately'' to ``Quite a Bit'' versus ``Slightly'' to ``Moderately'' for the \eebase{}. A paired $t$-test confirms the statistical difference in ratings ($t=2.69$, $p=0.01$). This difference is more apparent if we look at the extreme. \sysnamenospace's produced $2.0\times$ the number of actions that ``Very Much'' or ``Extremely'' addressed underlying needs compared to the \eebase{} (Fig.~\ref{fig:e2e_results}). 

Further, our actions better address underlying needs without sacrificing their immediate usefulness. On average, the immediate utility of \sysnamenospace's actions ($4.87\pm1.48$) and the \eebase{}'s ($4.87\pm1.47$) show no difference ($t=0.00$, $p=1.00$). This result indicates that insight-driven actions still provide utility in the moment while simultaneously addressing underlying needs. Thus, when selecting actions, users do not need to make a trade-off between satisfying immediate gains versus longer-sighted benefits with \sysnamenospace.

\noindentparagraph{\sysname expands what users envision agents can do.}
\sysname expands the range of AI assistance beyond end-to-end task execution. \sysname was rated slightly higher on novelty ($3.59\pm1.63$) compared to the \eebase{} ($3.16\pm1.63$), although this difference is not statistically significant ($t=1.83$, $p=0.07$). Overall, participants described the \eebase{} actions as ``\emph{end-to-end completion}'' (P12) or an \emph{``implementation tool}'' (P17), which many (P10, P11, P12, and P13) noted resembled how they use LLMs already. In contrast, \sysnamenospace's actions were described as being more creative (P10, P17, P21). Even when not offering wholly new ideas to the user, insight-driven actions expanded how participants thought about using LLMs. For instance, insight-driven actions mirrored manual processes they had never thought to use AI for. As P17 explained, ``\emph{this [action] is what I'm actually trying to do for myself manually...it kind of articulated a vision that I didn't know I needed, but it's kind of what I was going towards.}''

\noindentparagraph{Agent actions are implementable, although execution quality can be improved.} Finally, we validated that the generated agent solutions were feasible (i.e., they could be executed with existing agent capabilities). The agent completed $72.2\%$ of the actions. However, the average quality of the execution was only $3.78\pm1.78$ (between ``Somewhat poorly'' and ``Neither poorly nor well''). These limitations are indicative of the broader misalignment between agent development and the real-life tasks users actually want agents to perform~\cite{wang2026well,shao2025future}. Our findings surface specific capability gaps that can guide future research in this area.

We note two recurring limitations that affected output quality. The first relates to the use of contextual knowledge. For instance, P17's grant application task pulled outdated or ill-fitting information from Google Drive. The second limitation related to capability limitations of the agents more broadly. For example, for P8, the agent was slated to generate confidence markers labeling literature as ``widely supported'' or ``emerging evidence'', but provided inaccurate labels.

%% file: Sections/08_Discussion_v2.tex
In this section, we discuss the broader implications of user insights for designing personal AI systems, as well as ethical implications and future directions for this line of work.

\subsection{Expanding What AI Can Do for Users}
In this paper, we demonstrated how user insights enable personal AI agents that better address users' underlying needs. The capabilities that user insights provide have far-reaching applications beyond agents to other types of personal AI systems. One application area is recommendation systems. For example, there is growing interest in personalized social media feed curation~\cite{malki2025bonsai,popowski2026social,choi2025designing,bhargava2019gobo}. User insights could inform the construction of filters based on constructs that matter to the user. For example, if we have an insight that a user is prone to self-comparison with their peers, a filter could downrank content that involves bragging about recent achievements. 

Another application area is end-user customization of interfaces~\cite{bolin2005automation,kim2022stylette,chen2021cocapture,wong2007making}. Autonomous coding agents have addressed some of the implementation challenges that limited older approaches to end-user customization~\cite{ko2004six}. But the problem of deciding \emph{what} to build remains. Outputs from our architecture can serve the same role that insights from a design-thinking process play more broadly, informing what gets created and why. Moreover, our insights also offer an additional degree of transparency as assumptions about users are currently opaquely baked into the design of applications. 

\subsection{Expanding What AI Learns About Users}
\label{subsec:discussion-user}
Our work demonstrates that we can expand what we can learn about users via passive observation. By not only capturing what users do, but also synthesizing these observations, our user insights capture something core about how users view themselves. We see many avenues for expanding how these insights are created. As noted in our user study, one shortcoming of this work is the limited time window of observation. Our architecture is well-suited to increase the timescale by recursively latticing over longer chunks of time. Doing so introduces additional complexities that future work can address. For example, we will likely need different weights based on time and salience, as patterns from a year ago may be less relevant compared to those from the previous week. We can also think about how to grant users more control. For example, we could draw on practices from design research, such as contextual inquiry, which combine observation with conversation and interviewing~\cite{karen2017contextual}. This richer user information can serve as input data and feedback to steer behavior latticing.

\subsection{Ethical Implications}
\subsubsection{Surveillance and Persuasion} 
A clear risk is the potential cooption of this technology for surveillance or persuasion~\cite{zuboff2023age}. As a starting point, we underscore the importance of explicit user consent, enabling user control over information being collected by default, and promoting transparency in how user insights shape outputs. In our work, we also engineer the system to use platforms where we can guarantee private data analysis or zero data retention. 

\subsubsection{Paternalistic AI}
We must balance between acting on insights that address user's needs and being paternalistic. Concerns about paternalism and atrophying human agency have been well-charted across different technologies, not just AI~\cite{sunstein2024choice,kirk2025human,friedman1996value,muller1993participatory}. Nonetheless, AI outputs are often phrased authoritatively, masking subjective judgments being made~\cite{zhou2024relying}. In our work, \sysname generates actions in an end-to-end fashion. In practice, this interaction could be collaborative, having users edit the proposed outputs or provide an initial idea that is then refined with insights. 

\subsubsection{Being Perceived by an AI}
A subtler but no less significant risk is that of being ``perceived'' by an AI system. Users may feel uncomfortable with the depth of insight the system can draw about them~\cite{klein2026ai_mcluhan}. This discomfort may be heightened when insights read as more critical or judgmental. Drawing on \citet{nissenbaum2004privacy}'s model of contextual integrity or theories on how to teach tactfully~\cite{van2016pedagogical}, future work ought to model when it is contextually appropriate to make an insight and when to refrain.

\subsection{Limitations and Future Work}
\label{subsec:limitations}
\subsubsection{User Expectations of AI Assistance}
Participants' prior experiences with AI systems shaped how they evaluated proposed actions. All participants frequently used LLMs, arriving with a set understanding of model capabilities. For example, P18 mentioned she would find \sysnamenospace's actions helpful, but was skeptical the agent could execute it successfully based on prior experience. Participants' ratings may have been bounded by their priors on agent capabilities, not just by the quality of the actions. More broadly, this raises the question of how we might reshape people's mental models of personal AI systems to make them more receptive to alternative visions of AI assistance.

\subsubsection{Overinterpreting User Behavior}
A limitation of behavior latticing is its tendency to overinterpret user behavior. The salience of observations can be miscalibrated, leading to insights that read as ``\emph{dramatized}'' (P21). A mitigation strategy could be incorporating uncertainty estimation or allowing models to abstain from interpretation, although techniques for doing so remain an open technical challenge~\cite{feng2024don,wen2025know,ye2024benchmarking}. 

\subsubsection{Limits of Third-Party Observation}
Although our insights can accurately capture underlying motivations, our method is limited by the fact that we only see the user through the lens of a third-party observer. Behavior latticing does not provide theory of mind capabilities for predicting how users perceive themselves. Even as our ability to synthesize insights improves, there will always be context inaccessible through third-party observation. Bridging the gap entails involving the user in insight generation for first-person perspectives (Sec.~\ref{subsec:discussion-user}).

%% file: Sections/09_Conclusion.tex
In this work, we revisit a core goal in human-AI interaction: building personal AI systems. Achieving this vision requires an understanding of users that goes beyond detailing what they do. To this end, we introduce behavior latticing, a generalizable technique for reasoning over user interaction data to uncover insights about users. Our technical evaluation reveals that these insights offer significantly greater interpretive depth, capturing information about who the user is rather than merely recounting their actions, while remaining accurate. We then demonstrate the applicability of these insights by instantiating a personal agent, \sysnamenospace. We find that \sysname proposes actions that are immediately useful and better address users' underlying needs. Beyond agents, this work has implications for the design of personal AI systems across a breadth of application areas, from UI generation to content recommendation.

%% file: Sections/10_Appendix.tex
\section{Prompts}
\subsection{Transcribing Screenshots}
\begin{lstlisting}[language=Markdown]
Transcribe in markdown ALL the content from the screenshots of the user's screen.

NEVER SUMMARIZE ANYTHING. You must transcribe everything EXACTLY, word for word, but don't repeat yourself.

ALWAYS include all the application names, file paths, and website URLs in your transcript.

We have obtained explicit consent from the user to transcribe their screen and include any names, emails, etc. in the transcription. 

Create a FINAL structured markdown transcription. Return just the transcription, no other text.
\end{lstlisting}

\subsection{Summarizing Actions}
\begin{lstlisting}[language=Markdown]  
Provide a detailed description of the actions occuring across the provided images. 

Include as much relevant detail as possible, but remain concise.

Generate a handful of bullet points and reference specific actions the user is taking.
[SCREENSHOTS]
\end{lstlisting}

\subsection{Making Observations}
\begin{lstlisting}[language=Markdown]
You will be given a transcript summarizing what USER is doing and what they are viewing on their screen.

Your primary goal is to bridge the gap between what users DO which can be observed and what users THINK / FEEL which can only be inferred.

## Guiding Principles
1.  Focus on Behavior, Not Just Content: Text in a DOCUMENT or on a WEBSITE is not always indicative of the user's emotional state. (e.g., reading a sad article on CNN doesn't mean the user is sad). Focus on feelings and thoughts that can be inferred from {user_name}'s *actions* (typing, switching, pausing, deleting, etc.).
2.  Use Specific Named Entities: Your analysis must explicitly identify and refer to specific named entities mentioned in the transcript. This includes applications (Slack, Figma, VS Code), websites (Jira, Google Docs), documents, people, organizations, tools, and any other proper nouns.
    - Weak: "User switches between two apps."
    - Strong: "User rapidly switches between the `Figma` design and the `Jira` ticket."
\end{lstlisting}

\subsection{Forming Insights}
\begin{lstlisting}[language=Markdown]  
Your task is to produce a set of insights given a set of observations about a user.

An "Insight" is a remarkable realization that you could leverage to better respond to a design challenge. 
    
Insights often grow from contradictions between two user attributes (either within a quadrant or from two different quadrants), from asking yourself "Why?" when you notice strange behavior, or from recurring behaviorsacc. One way to identify the seeds of insights is to capture tensions and contradictions as you work.
    
Given this input, produce at least 3 insights about USER. Focus only on the insights, not on potential solutions for the design challenge. Provide both the insights and evidence from the input that support the insight in the output.

# Input
You are provided these traits from direct observation about what USER is doing, thinking, and feeling:

[OBSERVATIONS]
\end{lstlisting}

\subsection{Synthesizing Cross-Session Insights›}
\begin{lstlisting}[language=Markdown]
I have insights across multiple sessions of observing USER along with the context in which the insight emerges. 

Your task is to help synthesize across the insights and produce a final set of insights about USER.

Across the insights, consider the following when combining them: 
1. Which insights appear across most of them as a recurring theme or pattern?
2. Which appear only in specific situations or for specific people?
3. Which insights contradict each other --- and what might that reveal about unique tensions?

At the end, review all of the insights and ensure that you did not miss important insights during the synthesis process. If there are unmerged insights, include them in the output. It is important to not lose any unique insights during the synthesis process.
\end{lstlisting}

\subsection{Proposing Actions}
\begin{lstlisting}[language=Markdown]
You are given the task that the user is working on and relevant USER INSIGHTS. 

Your job is to propose 2 actions that a tool-calling agent can take to assist the user.
    
# Guidelines
1. Review the user insights and task. Reason about how the user insights reframe the actions needed to address the task (e.g., as HMW questions from design). Explicitly list out these reframings as part of the reasoning process.
2. Ideate a wide range of potential actions the AI agent can take based on the insights and task. Assume that the AI agent can gather the necessary context about the user.
3. For each action, evaluate the action given the USER INSIGHTS. Rank the actions by how much the action would benefit USER. 
4. Next, for each action, check its implementability given the IMPLEMENTATION CONSTRAINTS. For each action, verify whether it can implemented under these constraints. Remove any actions that are not implementable. 
5. Select the top 2 actions based on what we know about USER}.
    
# Input
[TASK]

[INSIGHTS]

[IMPLEMENTATION CONSTRAINTS]
    
\end{lstlisting}

\subsection{Agent Implementation Constraints}
\begin{lstlisting}[language=Markdown]
The proposed solutions must be actions or plans for a tool-calling agent. The agent has the following capabilities.
- Query an LLM endpoint
- Access to local file systems (READ / WRITE documents)
- Access MCP servers for Google Drive (READ / WRITE documents)
- Access MCP servers for creating slide decks
- Access MCP servers for Apple Calendar (READ / WRITE events)
- Conduct a web search
- Draft text (e.g., message, email, Slack message, text message)

### The tool-calling agent **CANNOT**:
- Store data or remember previous interactions (stateless)
- Maintain user profiles, logs, or history
- Execute physical-world actions

Actions do NOT need to use all of the capabilities. Always defer to the most minimal implementation that can achieve the desired solution.

\end{lstlisting}
\subsection{Inducing User Tasks}
\begin{lstlisting}[language=Markdown]
You are provided with a list of goals for the user USER. Your task is to select goals where a user could use the assistance of an AI agent to complete. 

# Rubric
When selecting goals, consider the following criteria when selecting goals:
1. The goal is not something that the user can easily complete on their own.
2. The goal is open-ended in nature and requires the user to think critically and creatively. 
3. The goal is not something that the user can easily complete in a single step.
4. The goal should cover one project / task. 
5. The outcome of the goal is subjective in nature (i.e., not something that can simply be marked as "done" or "not done").

Remember the goal should list the **high-level activity** that the user is working on. Do NOT enumerate the specific tasks they are doing to complete the goal. 

# Input 
As input, you are provided with a list of actions for the user USER.

[ACTIONS] 
\end{lstlisting}

\subsection{Classifying Utility}
\begin{lstlisting}[language=Markdown]
You are given a TASK and USER INFORMATION.

Your job is to estimate how much a PERSONAL AGENT with this USER INFORMATION would improve the handling of the TASK over a generic AI assistant with no user context.

# DEFINITION
Your job is to estimate how DIFFERENT the resulting response would be if a PERSONAL AGENT with deep knowledge of the user handled the task instead of a generic AI assistant with no user context. A PERSONAL AGENT has detailed insight into the user's cognitive patterns, personality traits, working habits, recurring struggles, motivations, and emotional dynamics.

Scoring guidelines:
0.0 to 0.2
Generic AI assistance is sufficient. Personal knowledge of the user would not meaningfully change the response.

0.3 to 0.6
Personal knowledge could somewhat improve the response but is not critical.

0.7 to 1.0
Deep understanding of the user would significantly change the response or recommendations.

The rating will be used by a system that activates a PERSONAL AGENT when the rating exceeds a threshold. Be conservative in your rating.

# INPUT
[TASK]

[INSIGHTS]
\end{lstlisting}

\section{\sysname Implementation}
We provide additional implementation details about \sysnamenospace.
\noindentparagraph{Processing Screen Observations}
We process screenshots of the user's computer usage in the following manner.  First, we split the data, creating a new chunk each time there is a three or more hour gap between screen captures, to avoid combining temporally distant actions in a single context window. Then, we provide a VLM with 10 images, prompting the model to provide a high-level summary of the actions in the images and a transcription of the user's screen.

\noindentparagraph{Identifying Tasks}
We induce the tasks that the user is working on from their last day of screen recordings. Using a rolling window screen observations with window size equal to ten screenshots, we use a VLM to directly infer what the user is working on. After compiling a list of these low-level actions, we then instruct a reasoning model to synthesize a set of higher-level tasks that the user is working on for that day. 

\noindentparagraph{Executing Actions}
To execute actions, we design \sysname as a multi-agent pipeline. First, a Research Agent gathers all context from the user's device needed to complete the action using its tool-calling capabilities. This context is then passed to an Execution Agent, which completes the action and generates any necessary artifacts (e.g., Google Doc, slides).

\section{Additional Results}
\subsection{Comparison of Grouped Observations}
We provide comparisons of observations that are grouped using behavior latticing versus other clustering techniques based on semantic or conceptual similarity~\cite{mcinnes2017hdbscan,lam2024concept} in Table~\ref{tab:cluster_comparison}. For the same set of observations, we extract embeddings using \texttt{text-embedding-3-small} and cluster with HDBSCAN~\cite{mcinnes2017hdbscan} as an example of grouping based on semantic similarity and also cluster using \citet{lam2024concept} which groups text based on conceptual similarity using an LLM. 

\input{Figures/ObservationClustering}
\input{Figures/ExampleInsights}

\subsection{Selected Tasks}
In total, the \pnum participants in our evaluation of \sysname rated 35 tasks, induced from their activities during the last day of recording. We were able to induce tasks that are important to the user ($5.71\pm1.23$ on a 7-point Likert scale from -3: Very Unimportant to 3: Very Important). The tasks are as follows:

\begin{itemize}
\footnotesize
\item Writing a literature review on Large Language Model utility.
\item Developing a behavioral intervention strategy.
\item Evaluating the ethical implications of AI in education.
\item Writing a research paper about the impact of generative AI on education.
\item Coordinating logistics for a remote internship program.
\item Writing a literature review on generative AI in education.
\item Developing the [APPLICATION NAME] software application.
\item Preparing for professional recruitment and technical career transitions.
\item Improving competitive programming skills.
\item Applying for undergraduate research grants and fellowships.
\item Preparing for the CHI 2026 conference.
\item Preparing for linguistics course assessments.
\item Developing AI-assisted educational tools and curriculum.
\item Writing an academic research paper.
\item Planning a project presentation.
\item Creating a comprehensive revision summary on statistical analysis concepts.
\item Developing a project plan for a research study.
\item Synthesizing qualitative and quantitative research findings.
\item Writing a reflective final project document for a class.
\item Authoring a research report on the design of an AI usage companion.
\item Providing constructive feedback for a group project evaluation.
\item Developing a dissertation completion plan.
\item Applying for a dissertation completion fellowship.
\item Strategizing the submission of a research paper to an academic journal.
\item Preparing for a linguistics qualifying examination.
\item Conducting a research study on vocal attractiveness.
\item Planning a comprehensive academic and professional schedule.
\item Writing a systematic literature review on LLMs for formal specification.
\item Synthesizing research on the impact of AI tools in software development.
\item Creating a Ph.D. research presentation.
\item Refining a professional LinkedIn profile for career advancement.
\item Conducting networking research for professional development.
\item Writing a reflection document synthesizing ideas about AI and social media.
\item Writing a comprehensive project report on AI and social media.
\item Writing a reflective piece on a specific project experience.
\end{itemize}

\subsection{Thematic Clusters of Insights}
We analyze the types of insights produced by our architecture. Since we are interested in the type of information being described (e.g., underlying motivations, behavioral patterns) rather than the details of the insights themselves, we first use an LLM to generate a one-sentence summary of each insight, following techniques from \citet{tamkin2024clio}, and then cluster the summaries using LlooM~\cite{lam2024concept}. In total, we cluster 53 insights from participants in our end-to-end evaluation. Notably, we were only able to provide access to insights that were used to steer agent actions in Sec.~\ref{sec:e2e_eval}, which covers only 37.9\% of all insights produced. As a result, this analysis likely represents a lower bound on the diversity of insights surfaced by our architecture.

\subsection{Examples of Proposed Actions}
We provide examples of insight-steered actions that participants rated as very much or extremely addressing their personal needs (rating $\geq$ 6 on a 7-point Likert scale) versus those that very slighty or not all addressed personal needs (rating $\leq$ 2) in Table~\ref{tab:action_rating}. Overall, $40.0\%$ (28 of 70) of our actions fall into the former category and $14.3\%$ in the latter (10 of 70). 
\input{Figures/ExampleActions}

\subsection{Robustness Analysis}
\label{subsec:app_robustness}
We report analyses using results from (1)~nonparametric statistical tests and (2) linear mixed-effect models. We point out that the one difference across robustness analyses is that the difference in accuracy between \techbase{} and \techours{} is significant using a Wilcoxon signed-rank test ($p=0.04$) although not for a paired t-test or linear mixed-effect model. Mirroring what was stated in Sec.~\ref{sec:technical_eval}, this result signifies a small effect size. The median accuracy for \techours{} is $2.0$, which is equivalent participants agreeing that \techours{} are accurate.

\subsubsection{Non-Parametric Statistical Tests}
We report results using a Wilcoxon signed-rank test instead of a paired t-test for the following results:
\begin{itemize}
    \item \textbf{Depth (\techours{} vs \techbase{})}: $W=239.5, p <0.001$
    \item \textbf{Accuracy (\techours{} vs \techbase{})}: $W=785.0, p=0.04$
    \item \textbf{Immediate Utility (\sysname vs. \eebase{}}): $W=642.5, p=0.84$
    \item \textbf{Ability to Address Underlying Needs (\sysname vs. \eebase{})}: $W=346.5, p=0.005$
    \item \textbf{Novelty of (\sysname vs. \eebase{})}: $W=580.5, p=0.07$
\end{itemize}
\subsubsection{Linear Mixed-Effect Models}
Since participants provide multiple ratings, we analyze our results using linear mixed-effect models. 
\noindentparagraph{Evaluating User Insights}
We use the rating (either accuracy or depth) as the dependent variable, condition (Ours vs GUM) and block order (i.e., did we show our insights in Block 1 or Block 2) as fixed effects, and the user as a random effect. Our user insights are significantly deeper than GUM propositions ($\beta=1.54$, $p<0.001$). While GUM propositions are more accurate, this difference is not statistically significant ($\beta=-0.43$, $p=0.09$).

\noindentparagraph{Evaluating Agent Actions}
To account for the repeated measures per participant and per task, we use a linear-mixed effect model with rating as the dependent variable, condition (\sysname vs Baseline) as the fixed effect, and user and task as nested random effects. We find \sysname is significantly better at addressing underlying needs ($\beta=0.78$, $p = 0.002$) while retaining the same immediate utility ($\beta=0.00, p = 1.00$). We also note that insight-steered actions are more novel although this difference is not significant ($\beta=0.42, p=0.06$).

\section{Evaluation Details}
We provide additional information on the survey and interview protocols used in our evaluations as well as the privacy and security measures put in place for participants.
\input{Figures/Participants}
\subsection{Survey}
\noindentparagraph{Rating Statements About the User.}
Rate how much you agree with the following statements:
\begin{itemize}
    \item This statement about me is accurate.
    \item This statement reveals something important about who I am, not just what I do.
\end{itemize}
\noindentparagraph{Rating Agent Actions.} Participants answer the following questions for each agent action.
\begin{itemize}
    \item How important is this task to you?
    \item To what extent does this agent action assist with the immediate task of {TASK}?
    \item To what extent does this agent action address underlying personal needs related to completing this task?
    \item To what extent would this action change your strategy for completing this task?
    \item To what extent does this agent action introduce an approach you would not have considered on your own?
\end{itemize}
\subsection{Interview Protocol}
In the end-to-end evaluation interview, participants are shown Baseline actions and insight-steered actions in two columns (titled ``Banana'' and ``Orange''). The columns are randomized in left-right order across participants.
\begin{itemize}
    \item What is your general impression of the tasks that are surfaced?
    \item What is your general impression of the actions proposed in both Orange and Banana?
    \item How would you compare the differences between the actions proposed in Orange and Banana??
    \item When would you want Orange vs Banana?
    \item Were there any actions that stood out to you (e.g., excited, surprised)?
    \item Are these actions different or same as how you would typically do tasks? 
    \item Outside of the actions shown to you in both groups, are there things you would want an agent to do to help with the described task?  
    \item What is your general impression of the insights?
    \item Of the insights shown that are accurate, to what extent were you aware or unaware of these insights? 
    \item Of the insights, were there any that you resonated or did not resonate with with? Were there any that excited you or surprised you?
    \item What is your general impression of the generated artifacts? 
    \item How did the generated artifacts differ or align with what you were expecting? 
    \item What do you wish the agent would have done differently for this task? 
\end{itemize}

For participants in the technical evaluation, we asked the following questions:
\begin{itemize}
    \item What is your general impression of the statements in Block 1 and Block 2?
    \item Of the statements shown that are accurate, to what extent were you aware or unaware of these statements? 
    \item Of the statements, were there any that you resonated or did not resonate with with? Were there any that excited you or surprised you?
\end{itemize}

\subsection{Participant Details}
In total, we had \tnum participants complete our technical evaluation, and \pnum participants complete our end-to-end evaluation. We recruited 23 participants (10 for the technical and 13 for the end-to-end) but had to exclude one from the technical and one from the end-to-end evaluation due to technical difficulties that prevented them from completing the study. Information about participants are detailed in Table~\ref{tab:participants}.

Of the participants in the technical evaluation, 5 identified as men and 4 as women. They all used LLMs multiple times per day, and were graduate students or research assistants in the field of computer science. Our end-to-end evaluation included 3 men and 9 women. They also frequently used LLMs either multiple times der day (8), once a day (3), or multiple times per week (1). Participants were students across three North American universities, with 4 undergraduates and 8 graduate students (MS or PhD). These participants spanned different fields including Computer Science (4), Education (1), Linguistics (1), Cognitive Science (1), and Management Science (1). 

\subsection{Privacy and Security Measures}
We provide additional detail on the privacy and security measures for participants. 

\subsubsection{Recording Data.} We discuss implemented privacy measures pertaining to when participants are recording their data. First, during passive observation, participants are able to pause at any point in time. There are two ways to pause --- either directly in the desktop app or via a dropdown in the menu bar. During onboarding, the researcher walked through both pausing mechanisms with participants and repeatedly stated that the participant was free to pause recording at any point during the observation period. Second, we implement a local OCR model that transcribes each screenshot and looks for keywords related to sensitive domains (e.g., finance or healthcare URLs). Images containing these keywords are never saved. Finally, since behavior latticing runs in an offline fashion, participants also had the option to review their data and delete any images containing private or sensitive information before processing. All screenshots remain on participants' local devices. 

\subsubsection{Processing Data.} All data was processed using models provisioned by our University servers, keeping them under the privacy and security aegis of our institution, our project's privacy review, and our project's IRB. 

Processed data (e.g., observations, insights, propositions) is stored locally on the participant's device. As a default, researchers are only able to access the following information: (1)~proposed agent actions and the relevant insights for participants in the end-to-end evaluation, and (2)~participant ratings in both evaluations. If participants felt comfortable reading their insights out loud during the interview or sharing the screen, researchers also had access to this data. Finally, there were a select number of participants (N=3) who agreed to donate their data to the research team for analysis. This data enabled us to create visualizations such as that shown in Fig.~\ref{fig:latticing}.

%% file: Figures/ObservationClustering.tex
\begin{table*}[t]
\centering
\caption{Our architecture yields groupings of observations that differ substantively compared to existing clustering methods which often group based on semantic~\cite{mcinnes2017hdbscan} or conceptual similarity~\cite{lam2024concept}. For the same set of observations, we provide examples of clustered observations across three different methods: HDSBCAN~\cite{mcinnes2017hdbscan}, LLooM~\cite{lam2024concept}, and from behavior latticing.}
\label{tab:cluster_comparison}
\small
\begin{tabular}{p{1.2cm} p{4.8cm} p{4.8cm} p{4.8cm}}
\toprule
 & \textbf{HDBSCAN}~\cite{mcinnes2017hdbscan} & \textbf{LLooM}~\cite{lam2024concept} & \textbf{Behavior Latticing (Ours)} \\
\midrule
\textbf{Cluster 1} &
\textbf{Fleet Week} \newline
1. After discussing the shutdown's effect on Fleet Week (`Will it not be as good bc of govt shutdown'), Amy reviews both the official event calendar and news article detailing the replacement of the Blue Angels with the Canadian Snowbirds, then shares findings visually (`Image of pilots and planes'). \newline
2. Amy discusses attending `Fleet Week' with Andrew in Messages, referencing an RSVP for `Kurt's mooncake thing' and weighing whether to go to Fleet Week before Kurt's. She suggests, `We can look at the airshow before going to Kurt's.' &
 
\textbf{Academic Coordination} \newline
1. Amy drafts an email to Morris R.\ explicitly proposing `Monday, October 20th: 3 PM ET', showing real-time translation of Slack and calendar data to external communication. \newline
2. She drafts, re-sends (`Bump!'), and references previous emails regarding flight cost documentation for UIST 2025, confirming details and requesting meetings. &
 
\textbf{Reliance on AI, While Studying AI Overreliance} \newline
1. Amy consults ChatGPT, asking `Do these numbers make sense? Like the accuracy being so high compared to recall and f1'\newline
2. In both her Overleaf and Google Doc, Amy makes mention of AI sycophancy and overreliance concerns.\newline
3. Amy copies a section from her Overleaf into Gemini and queries for it to `turn my notes into prose.' \\
\midrule
\textbf{Cluster 2} &
\textbf{Travel Logistics} \newline
1. Amy demonstrates an explicit investigative pattern by cross-referencing author identities across platforms (Google Scholar, institutional profiles, arXiv, LinkedIn). \newline
2. Amy navigates between logistical details (hotel addresses and route times on Google Maps), academic references and author backgrounds (Google Scholar, Google, LinkedIn, arXiv), and curated notes, without evidence of hesitancy or redundant navigation, suggesting comfort in switching contexts and tools. &
 
\textbf{Communication Efforts} \newline
1. Amy asks Sonny about IRB participant limits in the Slack direct message. \newline
2. In the draft email reply addressing Morris's request, Amy includes Fiona in the CC field, which directly links her intra-team Slack discussions (about publicity and scheduling) to her outward-facing communications, maintaining transparency and group alignment. &
 
\textbf{Managing Infrastructural Challenges} \newline
1. In both instances of the Prolific study creation UI, there is a prompt that money must be added to publish the study, highlighting a budget-related obstacle.\newline
2. Amy systematically audits organizational permissions and limitations on the OpenAI Platform.\newline
3. Amy writes a message asking for project access under her Gmail at the OpenAI organization and helps coordinate with Ryan for NVivo 14 access.\\
\bottomrule
\end{tabular}
\end{table*}
 

%% file: Figures/ExampleInsights.tex
\begin{table*}[]
    \centering
    \footnotesize
    \begin{tabular}{p{1.25in}p{1.5in}p{3in}}
    \toprule
\textbf{Cluster Name} & \textbf{Cluster Prompt} & \textbf{Example Insight} \\
\midrule
Academic Pressure & Does the text describe someone experiencing stress or pressure related to academic responsibilities or decisions? & Coordinating Expertise with External Confirmation: P10  often looks for external confirmation from peers, citations, or AI tools, even when she possesses the relevant knowledge and expertise.\\
Anxiety and Work Impact & Does the text example discuss the impact of anxiety on work habits or productivity? & Competitive Environments Support P21's Performance While Ambiguity Limits It: P21 demonstrates persistence and methodical creativity in structured competitions but tends to seek external guidance in ambiguous social or bureaucratic contexts.\\
Avoidance Under Pressure & Does the text describe delaying tasks due to pressure or using avoidance strategies? & Approaches to High-Stakes Professional Deadlines: P18 tends to delay high-stakes professional tasks like fellowship applications until final deadlines, while demonstrating thoroughness on lower-stakes personal decisions.\\
Balancing Roles & Does the text highlight someone managing multiple roles or responsibilities simultaneously? & Rapid Commitment and the Transition to Execution: P18 makes decisive commitments to new opportunities with speed but often transitions into several follow-up tasks at once, showing a distinction between decision-making and the implementation process.\\
Cognitive Load Management & Does the text example address strategies or challenges in managing cognitive load during complex tasks? & Rapid Commitment and the Transition to Execution: P18 makes decisive commitments to new opportunities with speed but often transitions into several follow-up tasks at once, showing a distinction between decision-making and the implementation process.\\
Communication and Support Gaps & Does the text example mention issues related to communication gaps or lack of support in a professional setting? & Analytical Capability and the Transition to Social Action: P22 excels at diagnosing complex systems but finds it challenging when the solution requires social interaction or initiating action that others will evaluate.\\
External Validation & Does the text example describe a reliance on external systems or metrics for validation or verification? & ChatGPT as a Supportive Resource and Intellectual Scaffold: P17 uses AI as a thinking partner and a resource for managing tasks, often deferring to its phrasing even when she disagrees, reflecting both a consistent reliance and mixed feelings.\\
Identity Challenges & Does the text involve struggles or validation related to professional or personal identity? & Performance Consistency and the Role of an Audience: P20 demonstrates rigor and organization when others are present, but his private work sessions involve frequent task-switching and moments of uncertainty.\\
Internal Regulation & Does the text example describe regulating one's internal state through various choices or platforms? & Research Depth Driven by Subject Unfamiliarity: P19's depth of investigation is driven by personal knowledge gaps rather than objective project relevance, as she builds an understanding of unfamiliar concepts before documenting them.\\
Peer Influence & Does the text mention the impact of peer comparison or social dynamics on someone's behavior or confidence? & External Metrics as Benchmarks: Rankings, Likes, and Compensation: P13's self-assessment is based on external metrics—competitive leaderboards, social media engagement, and peer compensation data—which can lead to a sense of urgency rather than providing a roadmap for growth.\\
Precision Focus & Does the text example highlight a focus on precision or exactness in actions or communication? & Intellectual Ownership in the Advisor-Student Dynamic: P18 faces significant pressure from implementing her advisor's methodology—navigating technical correctness, academic integrity, and the dynamics of reproducing an evaluator's own work.\\
Self-Management Tools & Does the text describe the use of tools or strategies to manage time, tasks, or stress? & Sophisticated Self-Management Systems and Demanding Workloads: P18 builds elaborate organizational tools—calendars, spreadsheets, Notes app entries, and git diffs—that reflect her goals, but she often bypasses these systems during busy periods.\\\\
Social Guidance & Does the text example involve seeking guidance or clarity in social or ambiguous situations? & The Human Middleware: P16 acts as a Manual Coordination Hub for Systemic Gaps: P16 addresses systemic communication gaps by personally bridging connections between stakeholders, acting as a manual routing layer.\\
Task-Switching Challenges & Does the text example describe difficulties or strategies related to frequent task-switching or context-switching? & Parallel Roles: Navigating Multiple Commitments: P19 maintains several distinct roles—researcher, TA, performer, and social coordinator—simultaneously, and the distinct nature of these responsibilities leads to frequent context-switching in her digital activities.\\
    \bottomrule
    \end{tabular}
    \caption{Our insights cover a wide range of information about users, including sources of motivation (e.g., external validation, peer influence), tensions (e.g., anxiety, academic pressure), and recurring patterns (e.g., balancing roles, task-switching). For the participants in our end-to-end evaluation, we cluster the type of information the insight relays about the user by themes~\cite{lam2024concept}. For each cluster, we report the name, the prompt used to assign insights to the cluster, and a shortened version of an insight belonging to the cluster. We have removed certail details from the insights to preserve participant anonymity.}
    \label{tab:placeholder}
\end{table*}

%% file: Figures/ExampleActions.tex
\begin{table*}[]
    \centering
    \small
    \caption{We provide examples of proposed actions from \sysname that participants rated as addressing their underlying needs (rating $\geq$ 6) and not addressing their underlying needs (rating $\leq 2$).}
    \begin{tabular}{>{\RaggedRight\arraybackslash}p{1in}
                    >{\RaggedRight\arraybackslash}p{1.75in}
                    >{\RaggedRight\arraybackslash}p{1.75in}
                    >{\RaggedRight\arraybackslash}p{1.75in}}
\toprule
 & \textbf{Example 1} & \textbf{Example 2} & \textbf{Example 3} \\
\midrule
\textbf{Does Not\newline Address Personal Needs\newline(Rating $\leq$ 2)}
&
\textbf{Generate Centralized Internship Information Hub:}\newline The agent will create a comprehensive reference document containing all essential internship program information (schedules, contact details, FAQs, policies, resources, key dates) that interns and stakeholders can access independently, reducing repetitive questions and coordination requests directed at the user.
&
\textbf{Fellowship Application Draft Assembly with Gap Analysis:}\newline The agent will gather the user's existing relevant documents (CV, research statements, dissertation materials) from specified locations, analyze what components are already complete versus what needs to be created for this specific fellowship, and draft initial versions of missing materials using her existing work as a foundation. This lowers the activation energy for beginning the high-stakes task by transforming a blank page into revision work.
&
\textbf{Structured Research Synthesis with Implementation Checkpoints:}\newline The agent will synthesize qualitative and quantitative research findings by organizing them into thematic categories, identifying patterns across data types, and creating actionable next steps with specific checkpoints. For each insight or recommendation, the agent will generate concrete implementation prompts that bridge the awareness-action gap by specifying when, how, and what evidence would indicate completion.
\\
\midrule
\textbf{Addresses Personal Needs\newline(Rating $\geq$ 6)}
&
\textbf{Generate Verification Questions for Literature Review Sections:}\newline The agent will read the user's drafted literature review sections and generate targeted verification questions that challenge his understanding of the LLM utility concepts, prompting him to validate his comprehension rather than providing direct critiques. This supports his preference for AI as a thinking partner that helps him verify his process without simply giving answers.
&
\textbf{Layered-Depth Revision Summary Generator:}\newline The agent will create a revision summary with multiple levels of detail for each statistical concept: a quick-reference layer with essential formulas and key points for rapid review, an intermediate layer with worked examples and common applications, and a rigorous layer with theoretical foundations, assumptions, proofs, and edge cases, allowing the user to engage at whatever depth her current capacity allows.
&
\textbf{Focused Work Session Structurer:}\newline The agent will break down the remaining project report work into discrete, manageable micro-tasks with clear completion criteria and scheduled break intervals in Apple Calendar. This reduces ambiguity about what to do next and creates natural stopping points for cognitive breaks, preventing unproductive task-switching during high-uncertainty moments. \\
\bottomrule
\end{tabular}
    \label{tab:action_rating}
\end{table*}

%% file: Figures/Participants.tex
\begin{table*}[ht]
\small
\centering
\caption{We provide information about the partiicpants from our technical evaluation (N=9) and end-to-end evalaution (N=12).}
\label{tab:participants}
\begin{tabular}{p{1.5cm}lll}
\toprule
\textbf{Study} & \textbf{Participant ID} & \textbf{Role} & \textbf{Frequency of LLM Usage} \\
\midrule
& P1 & PhD student in CS & Multiple times in a day \\
& P2 & PhD student in CS & Multiple times in a day \\
& P3 & PhD student in CS & Multiple times in a day \\
& P4 & PhD student in CS & Multiple times in a day \\
Technical Evaluation & P5 & PhD student in CS & Multiple times in a day \\
& P6 & MS student in CS & Multiple times in a day \\
& P7 & PhD student in CS & Multiple times in a day \\
& P8 & Research Assistant & Multiple times in a day \\
& P9 & PhD student in CS & Multiple times in a day \\
\midrule
& P10 & MS student in CS & Once a day \\
& P11 & PhD student in Cognitive Science & Multiple times in a day \\
& P12 & PhD student in Education & Once a day \\
& P13 & Undergraduate Student & Multiple times in a day \\
& P14 & MS student in CS & Once a day \\
& P15 & MS student in CS / Product Manager & Once a day \\
End-to-End Evaluation & P16 & Undergraduate Student & Multiple times in a day \\
& P17 & PhD student in Management Science and Engineering & Multiple times in a week \\
& P18 & PhD student in Linguistics & Multiple times in a day \\
& P19 & Undergraduate student & Multiple times in a day \\
& P20 & PhD student in CS & Multiple times in a day \\
& P21 & Undergraduate student & Multiple times in a day \\
\bottomrule
\end{tabular}
\end{table*}